\documentclass[12pt,a4paper]{article}

\usepackage{amsmath,amssymb}
\usepackage[bookmarks=true]{hyperref}
\usepackage[nosort]{cite}
\usepackage[bulletsep]{collect}
\def\gfxon{\usepackage[final]{graphicx}}

\gfxon

\makeatletter
\let\old@startsection=\@startsection
\renewcommand{\@startsection}[6]{\old@startsection{#1}{#2}{#3}{#4}{#5}{#6\mathversion{bold}}}
\makeatother

\def\eeq{\end{eqnarray}}

\def\=:{=\hspace{-.7em}\raisebox{1.1ex}{.}\hspace{.1em}\raisebox{-0.2ex}{.} }

\newcommand{\beqn}{\begin{eqnarray}}
\newcommand{\eeqn}{\end{eqnarray}}

\newcommand {\beq}{\begin{eqnarray}}
\newcommand {\eeqq}{\end{eqnarray}}

\newcommand {\dagg}{^{\dagger}}

\newcommand {\lam}{\lambda}

\newcommand {\Tr}{{\rm Tr}\,}

\def\changed#1{ #1}

\makeatletter \@addtoreset{equation}{section} \makeatother

\makeatletter
\let\old@makecaption=\@makecaption
\def\@makecaption{\small\old@makecaption}
\makeatother

\makeatletter
\def\mr@ignsp#1 {\ifx\:#1\@empty\else #1\expandafter\mr@ignsp\fi}%
\newcommand{\multiref}[1]{\begingroup
\xdef\mr@no@sparg{\expandafter\mr@ignsp#1 \: }%
\def\mr@comma{}%
\@for\mr@refs:=\mr@no@sparg\do{\mr@comma\def\mr@comma{,}\ref{\mr@refs}}%
\endgroup}
\makeatother

\ifx\href\asklfhas\newcommand{\href}[2]{#2}\fi

\begin{document}

\begin{flushright}\footnotesize
\texttt{FTPI-MINN-10/37} \\
\texttt{UMN-TH-2931/11}
\vspace{0.5cm}
\end{flushright}
\vspace{0.3cm}

\renewcommand{\thefootnote}{\arabic{footnote}}
\setcounter{footnote}{0}
\begin{center}%
{\Large\textbf{\mathversion{bold}
Non-Abelian Monopoles in the Higgs Phase
}
\par}

\vspace{1cm}%

\textsc{Muneto Nitta$^{1}$ and  Walter Vinci$^{2}$}

\vspace{10mm}
$^1$\textit{Department of Physics, and Research and Education Center 
for Natural Sciences,\\ Keio University, 4-1-1 Hiyoshi, Yokohama, 
Kanagawa 223-8521, Japan}
\\
\vspace{.3cm}
$^2$\textit{University of Minnesota, School of Physics and Astronomy\\%
116 Church Street S.E. Minneapolis, MN 55455, USA}

\vspace{7mm}

\thispagestyle{empty}

\texttt{nitta(at)phys-h.keio.ac.jp}\\
\texttt{vinci(at)physics.umn.edu} 

\par\vspace{1cm}

\vfill

\textbf{Abstract}\vspace{5mm}

\begin{minipage}{12.7cm}

We use the moduli matrix approach to study the moduli space of 1/4 BPS kinks supported by vortices in the Higgs phase of  $\mathcal N=2$ supersymmetric $U(N)$ gauge theories when non-zero masses for the matter hypermultiplets are introduced. We focus on the case of degenerate masses. In these special cases vortices acquire new orientational degrees of freedom, and become ``non-Abelian''. Kinks acquire new degrees of freedom too, and we will refer to them as ``non-Abelian''. As already noticed  for the Abelian case, non-Abelian kinks must correspond to non-Abelian monopoles of the unbroken phase of $SU(N)$ Yang-Mills.

 We show, in some special cases, that the moduli spaces of the two objects are in one-to-one correspondence. We argue that the correspondence holds in the most general case. 
 
 The consequence of our result is two-fold. First, it gives an alternative way to construct non-Abelian monopoles, in addition to other well-known techniques (Nahm transform, spectral curves, rational maps). Second, it opens the way to the study of the quantum physics of non-Abelian monopoles, by considering the simpler non-Abelian kinks.

\end{minipage}

\vspace{3mm}

\vspace*{\fill}

\end{center}

\newpage

\section{Introduction}
The history of magnetic monopoles is quite long, and it tracks back to the successful  attempt of Dirac to introduce magnetic charges into a consistent quantum  mechanics of charged particles \cite{Dirac:1931kp}. Existence of monopoles was motivated by the explanation of the quantization of the electric charge. Another step toward the legitimization of monopoles as an important subject of study was made by  't Hooft \cite{'tHooft:1974qc} and Polyakov \cite{Polyakov:1974ek}, who showed that magnetic monopoles where  necessarily present in  many spontaneously broken gauge theories, including all theories of grand unification \cite{Preskill:1979zi,Zeldovich:1978wj}. While from the experimental point of view monopoles are still problematic, since they have never been observed in nature, from the theoretical point of view they have been inspiring successful ideas in various area of physics. In cosmology, for example, they motivated the introduction of the concept of inflation by {Sato and} Guth \cite{Sato:1980yn,Guth:1980zm}. Monopoles were also considered as playing a crucial role in the strongly coupled dynamics of gauge theories. In particular, condensation of monopoles in the vacuum of QCD  can explain confinement in terms of a dual superconductivity mechanism \cite{'tHooft:1981ht,Mandelstam:1974pi}.  At the same time, the study of particular monopole solutions, called BPS \cite{Bogomolny:1975de,Prasad:1975kr}, whose energy is proportional to the magnetic charge, was particularly fruitful. 
BPS monopoles naturally arise in theories with extended supersymmetry, where the proportionality of masses and the existence of degenerate solutions are explained in terms of central charges and unbroken supersymmetry. More recent results in the non-perturbative dynamics of $\mathcal N=2 $ supersymmetric theories have been used to prove the actual role of magnetic monopoles in the mechanism of confinement \cite{Seiberg:1994rs,Seiberg:1994aj}. 

The study of monopoles in theories with residual non-Abelian gauge symmetries has also motivated the idea of dualities as crucial properties of gauge theories. The first proposed example was a direct generalization of the electro-magnetic duality to the non-Abelian case  \cite{Goddard:1976qe,Montonen:1977sn}. The presence of electro-magnetic dualities, which also involve a weak-strong coupling duality, is now well established in the context of supersymmetric gauge theories and string theory. A few  examples are Seiberg duality in  $\mathcal N=1$ theories \cite{Seiberg:1994pq}, Seiberg-Witten duality in $\mathcal N=2$ \cite{Seiberg:1994rs,Seiberg:1994aj}, S-duality in $\mathcal N=4$ \cite{Sen:1994yi}. In all of these cases, quantum monopoles play a crucial role. Moreover,  $\mathcal N=2$ theories can support vacua where non-Abelian gauge symmetries are unbroken in the infrared \cite{Carlino:2000ff,Carlino:2000uk}. In this context, semiclassical configurations of non-Abelian monopoles \cite{Ward:1982ft,Weinberg:1982ev,Auzzi:2004if} play a crucial role. However, despite many efforts, the quantum nature of these objects is still quite mysterious. This is due to a number of reasons. For example, the residual non-Abelian dynamics is usually strongly coupled in the infrared, and semiclassical configurations cannot be trusted \emph{a-priori}. Even when one circumvents this problem by considering infrared free theories, the standard  semiclassical quantization of non-Abelian monopoles is still problematic \cite{Abouelsaood:1982dz,Abouelsaood:1983gw,Nelson:1983bu,Balachandran:1982gt,Horvathy:1984yg,Horvathy:1985bp,Horvathy:1988ge} despite various attempts \cite{Nelson:1983fn,Dorey:1995me,Dorey:1996jh,Bais:1997qy,Auzzi:2004if}.

Despite these difficulties, there is a large amount of literature devoted to the construction of the most general configuration of classical monopoles. In the BPS limit, the second order equations of motion \changed{are simplified} into the first order Bogomol'nyi equations, which  admit a large set of continuously connected solutions called moduli space. Unfortunately, the explicit construction of monopole solutions is, in general, very difficult. Despite the complicate nature of the problem,  a surprisingly large number of different approaches and auxiliary methods have been developed during the years to pursue this task. The Nahm transform \cite{Nahm}, which is a direct adaptation of the Atiya-Drinfeld-Hitchin-Manin construction for instantons \cite{Atiyah:1978ri}, and the spectral curve approach  \cite{Hitchin:1982gh} are two such examples. Other alternative methods are related to twistors \cite{Ward:1977ta} and integrable systems \cite{Forgacs:1980ym}. We will not discuss  these methods in detail, rather we will make use of  the  rational map approach \cite{Donaldson:1985id,Jarvis,Jarvis2,Dancer,Murray:1989zk}.

With the present paper we propose a different approach to study classical configurations of BPS non-Abelian monopoles. Following the ideas developed in Refs. \cite{Tong:2003pz} for the Abelian case, we put non-Abelian monopoles in the Higgs phase\footnote{The idea of considering non-Abelian monopoles in the Higgs phase is not new  \cite{Auzzi:2003fs}. The main advantage of our set-up is that  monopoles in the Higgs phase are still 1/4 BPS, while in the approach of Ref. \cite{Auzzi:2003fs} monopoles are confined by vortices in a non-BPS solitonic configuration. }.

The minimal set-up for the construction of non-Abelian monopoles is \changed{an} $SU(N)$ pure Yang-Mills with an additional adjoint field. By embedding this monopole theory into a larger model, with the inclusion of matter fields in the fundamental representation and the extension of the gauge group to $U(N)$, it is possible to track the fate of monopoles when we enter the Higgs phase. As  well-known, the magnetic flux in the Higgs phase will be confined by flux tubes. Monopoles, for the same reason, will necessarily become kinks connecting different vortices \cite{Hindmarsh:1985xc,Tong:2003pz}. This setup was successfully considered in Refs. \cite{Hanany:2004ea,Shifman:2004dr} to give a physical explanation of the correspondence between the BPS spectra of two and four-dimensional gauge theories \cite{Dorey:1998yh,Dorey:1999zk}. A strictly related fact is  the equivalence of the spectrum of classical excitations of Abelian monopoles and kinks \cite{Abraham:1992vb,Abraham:1992qv}.  Moreover, it was found that the moduli spaces of Abelian monopoles aligned on a line and of domain walls are isomorphic \cite{Hanany:2005bq,Eto:2008dm}. 

The main purpose of this paper is  the extension of  this correspondence of moduli spaces of monopoles and kinks in the most general case, including the non-Abelian.

  As a concrete mode, we consider $\mathcal N=2$ supersymmetric $U(N)$ SQCD with $N$ matter hypermultiplets. The theory has a Fayet-Iliopoulos (FI) parameter $\xi$ which can be turned on to put the theory into the Higgs phase. The symmetry breaking pattern is controlled by the values of the hypermultiplets bare masses. When two or more masses coincide, the vacuum of the theory has, at the classical level, unbroken, non-Abelian symmetries. For vanishing FI, the theory is in the unbroken phase, and these symmetries are local. When we turn on the FI, the surviving symmetries are global. As  we enter the Higgs phase, non-Abelian monopoles become ``non-Abelian'' kinks \cite{Eto:2008dm} interpolating between different type of  non-Abelian vortices. As a consequence,  we can identify the classical moduli space of non-Abelian monopoles with that of non-Abelian kinks. The construction of kinks is made in terms of the moduli matrix \cite{Eto:2006pg}. This allows us to propose a ``moduli matrix construction for monopoles''. We also notice a close similarity between moduli matrices and rational maps. This observation provides us with a  ``physical'' interpretation of the monopole rational map.
  
In section 2, we review the concept of non-Abelian monopoles and the determination of their moduli space in terms of rational maps. In section 3, we consider a supersymmetric theory which supports monopoles in the unbroken phase and a system of vortices and kinks in the Higgs phase. In section 4, we review the moduli matrix construction for kinks,  and apply it to study their moduli spaces. In section 5, we compare the moduli spaces of monopoles and kinks in two particular cases, explicitly showing their equivalence. Finally,  in section 6, we present and motivate a conjecture about the complete  correspondence between the moduli space of kinks and monopoles in the most general case.

\section{Non-Abelian Monopoles in the Unbroken Phase}
In this section we briefly review the concept of (non-Abelian) monopole as topological soliton supported by non-Abelian gauge theories. We also review the concept of moduli space, which is a crucial aspect in studying the dynamics of monopoles in the BPS saturated case.

\subsection{Monopole solutions in spontaneously broken gauge theory}
It is well-known, since the pioneering works of 't Hooft \cite{'tHooft:1974qc} and Polyakov \cite{Polyakov:1974ek}, that non-Abelian gauge theories with spontaneous symmetry breaking admit non-singular magnetic monopole solutions. The simplest example of such theories is pure $SU(N)$ Yang-Mills with an additional adjoint scalar field $\phi$:
\begin{eqnarray}
S & = & \int d^{4}x \left\{\frac1{4 g^{2}}(F^{a}_{\mu\nu})^{2}+\frac{1}{g^{2}} |D_{\mu}\phi^{a}|^{2} -\frac{\lambda}{4}\left( v-|\phi^{a}|^{2}  \right)^{2}\right\} \nonumber \\
  & = & \int d^{4}x \,\Tr \left\{\frac1{2  g^{2}}(F_{\mu\nu})^{2}+\frac{1}{g^{2}} |D_{\mu}\Phi|^{2} -\frac{\lambda}{2}\left( v/N-|\Phi|^{2}  \right)^{2}\right\}
\label{eq:Mon act}
\end{eqnarray}
with $|A|^2\equiv A\dagg A$ for a square matrix $A$.
In the second line, we have written fields as matrices in the following way:
\begin{eqnarray}
F_{\mu\nu}=F_{\mu\nu}^{a}\tau^{a}, \quad A_{\mu}=A^{a}_{\mu}\tau^{a}, \quad \Phi=\sqrt2\,\phi^{a}\tau^{a}, \nonumber \\
D_{\mu}\Phi =\partial_{\mu}\Phi-i[A_{\mu},\Phi], \quad{\rm Tr} (\tau^{a}\tau^{b})=\frac12\delta^{ab}.
\end{eqnarray}
The action contains a potential term which fixes the expectation value of the scalar field $\Phi$, triggering a spontaneous breaking of the gauge symmetry.
\beq
|\Phi_{0}|^{2}=v/N\,.
\label{eq:phizerovev}
\eeq 

We will be particularly interested in the Bogomol'nyi-Prasad-Sommerfeld (BPS) limit \cite{Bogomolny:1975de,Prasad:1975kr}: $\lambda \rightarrow 0$. In this case, a square root completion is possible \cite{Bogomolny:1975de}. For time independent configurations, the action (\ref{eq:Mon act}) reduces to an energy and can be rewritten as
\begin{eqnarray}
E & = & \int d^{3}x \, {\rm Tr} \left\{ \frac{1}{4 g^{2}}(\epsilon_{ijk}F_{jk}-D_{i}\Phi)^{2}+ \frac{1}{g^{2}}\partial_{i}(\epsilon_{ijk}\Phi F_{jk})     \right\}.
\end{eqnarray}
The energy is minimized when the first positively defined term is set to zero. This gives the Bogomol'nyi equations for BPS monopoles:
\begin{equation}
D_{i}\Phi=\epsilon_{ijk}F_{jk}.
\label{eq:bogom}
\end{equation}
The monopole mass is then given by the second term
\begin{equation}
M=\frac{1}{g^{2}} \int d^{3}x\,  {\rm Tr} \,\partial_{i}(\epsilon_{ijk}\Phi F_{jk}) =\frac{1}{g^{2}}\int d^{3}x \,\Tr \partial_{i}(B_{i}\Phi)  = \frac{1}{g^{2}} \int_{S^{2}_{\infty}} dS^{i} \, \Tr (B_{i}\Phi),
\label{eq:mass}
\end{equation}
where we have introduced the non-Abelian magnetic field $B_{i}\equiv\epsilon_{ijk}F_{jk}$.  The last term is proportional to the magnetic flux of the configuration, and, as we shall see later in more detail, is proportional to the topological invariants associated with the monopole. 

By an appropriate gauge transformation, the value of $\Phi$ at infinity can always be put into a diagonal form\footnote{In the BPS case, equation (\ref{eq:phizerovev}) doesn't exist, and the quantity  $v$ should be considered as a free boundary condition setting the energy scale for the monopole configuration.}:
\beq
\Phi_{\infty} = \left(
\begin{array}{ccc}
 \mu_{1} &  \dots & 0  \\
\vdots  &  \ddots & \vdots  \\
 0 & \dots &  \mu_{N}  
\end{array}
\right), \quad \sum_{i=1}^{N}\mu_{i}=0, \quad \mu_{i}\le\mu_{i+1}.
\label{eq:phimu}
\eeq
For generic values of the eigenvalues $\mu_{i}$, the gauge symmetry is maximally broken:
\beq
SU(N) \longrightarrow U_{1}(1)\times \dots\times U_{N-1}(1) ,
\label{eq:maximal}
\eeq
and monopoles are supported by the non-trivial homotopy group:
\beq
\pi_{2}\left( \frac{SU(N)}{U(1)^{N-1}}   \right)=\pi_{1}(U(1)^{N-1})=\mathbb Z^{N-1} .
\eeq
The topological sector is associated to $N-1$ integers $m_{i}$ which correspond to the $N-1$ magnetic charges   of the monopole. From a physical point of view, these integers arise as a direct consequence of the Dirac quantization condition for the magnetic charge \cite{Dirac:1931kp}.
In the maximally broken case {monopoles are called Abelian.}

In contrast, when two or more $\mu_{i}$'s are equal,  there is a surviving non-Abelian gauge symmetry:
\beq
SU(N)\longrightarrow S(U(n_{1})\times U(n_{2})\dots \times U(n_{q})),  \quad \sum_{i=1}^{q} n_{i}=N
\label{eq:nonmaximal}
\eeq
and monopoles are supported by the homotopy group:
\beq
\pi_{2}\left( \frac{SU(N)}{S(U(n_{1})\times U(n_{2})\dots \times U(n_{q}))}   \right)=\mathbb Z^{q-1}.
\eeq
{The case with $q=2$ with $SU(N)\longrightarrow SU(N-1)\times U(1)$ is called minimal breaking.}
Monopoles in the non-Abelian case were originally constructed as embeddings of the $SU(2)$ solutions \cite{Ward:1982ft,Weinberg:1982ev,Auzzi:2004if}. Non-trivial solutions which cannot be obtained as embeddings have been also  constructed as explicit solutions of the BPS equations \cite{Bais:1978yh,Weinberg:1982jh,Ward:1981qq}.

Let us now review a more formal approach to deal with monopoles in the most general case of symmetry breaking.  First, we can  write the expectation value of $\Phi$ in terms of a vector $\vec h$: 
\beq
\Phi_{\infty}\equiv  \vec h \cdot \vec H 
\label{eq:boundaryphi}
\eeq
{
with generators $\vec H$ in the Cartan subalgebra of ${\cal G}$.
}
The symmetry breaking pattern is determined by the alignment of $\vec h$ with respect to the roots   $\vec \beta_{i}$ of $SU(N)$. As  well-known from group theory, maximal symmetry breaking (\ref{eq:maximal}) is obtained when $\vec h$ has a non-zero projection along all of the roots
\beq
\vec h \cdot \vec \beta_{i}\neq0 \quad {\rm for \, each } \quad i=1,\dots,N-1,
\eeq
while a residual non-Abelian gauge symmetry remains unbroken as in (\ref{eq:nonmaximal}) whenever $\vec h$ is orthogonal to one ore more roots $\vec \beta_{i}$. 
By analyzing the Bogomol'nyi  equations (\ref{eq:bogom}) together with the boundary conditions (\ref{eq:boundaryphi}) it is possible to show that there  is a gauge choice such that the asymptotic behavior for the adjoint field is the following\cite{Goddard:1976qe,Weinberg:1979zt}:
\beq
\Phi\sim \Phi_{\infty}-\frac{G_{0}}{4 \pi \, r} +\mathcal O (\frac1{r^{2}})\,.
\label{eq:higgsboundary}
\eeq
As a consequence of the Bogomol'nyi equations, $\Phi_{\infty}$ and $G_{0}$ commute \cite{Weinberg:1979zt}, and, up to a gauge transformation, {both of them} can be chosen to  belong to the Cartan subalgebra. $G_{0}$ is then related to the asymptotic value of the magnetic field: 
\beq
B_{r}=\frac{G_{0}}{4 \pi \, r^{2}}+\mathcal O (\frac1{r^{3}})\,.
\label{eq:boundary}
\eeq
The matrix $G_{0}$, being an element of the Cartan subalgebra, 
\beq
G_{0}= \vec g \cdot \vec H\,,
\eeq
can be written in terms of a vector  $\vec m$  of magnetic charges 
as below. 
A crucial early result about non-Abelian monopoles is the existence of a generalized Dirac quantization condition \cite{Goddard:1976qe,Englert:1976ng,Montonen:1977sn}. It  states that the magnetic charge vector must belong to the dual root lattice of $SU(N)$:
\beq
\vec g=4 \pi\,\sum_{i}^{N-1} m_{i}\cdot \vec \beta_{i}^{*},
\eeq
where the integers $m_{i}$ are magnetic charges and  the dual roots  $\vec \beta_{i}^{*}$ are defined as usual:
\beq
\vec \beta_{i}^{*}=\frac{\vec \beta_{i}}{\vec \beta_{i}\cdot \vec \beta_{i}}.
\eeq
It is important to recognize that this quantization condition is valid regardless of the specific symmetry breaking pattern defined by $\vec h$. A consequence of this is that only a subset of the $N-1$ integers $m_{i}$ is related to a topological quantity \cite{Weinberg:1979zt}.  
{
First, $m_{i}$ can be divided into two classes
$\{m_i\}=\{m_{t}^{\rm top},m_{h}^{\rm hol}\}$ defined as follows.} The charge $m_{i}$ corresponds to a topological integer if  the root $\vec \beta_{i}$ defines a broken $SU(2)$ subgroup. In other words: 
\beq
m_{t}^{\rm top} : \quad   \vec \beta_{t}\cdot \vec h \neq0, \quad t=1,\dots,q-1.
\eeq
The remaining integers are called ``holomorphic'' charges and correspond to roots defining unbroken $SU(2)$ groups:\beq
m_{h}^{\rm hol} : \quad   \vec \beta_{h}\cdot \vec h =0, \quad h=1,\dots,N-q.
\eeq
Using the above formulas we can evaluate the expression for the monopole mass,
\beq
M_{\rm mon}= \frac{2\pi}{g^{2}} \sum^{N-1}_{i=1}m_{i}\,\vec h \cdot \vec \beta_{i}^{*}= \frac{2\pi}{g^{2}}\sum^{q-1}_{t=1}m_{t}^{\rm top}\,\vec h \cdot \vec \beta_{t}^{*},
\label{eq:monopolemassfinal}
\eeq
which depends, as expected, only on the topological charges.

As we will discuss in more detail in the next session, BPS monopoles come as a continuous family of degenerate solutions of the  equations of motion. This  moduli space is given by the disjoint union of sectors labeled by the topological charges. On the other hand, holomorphic charges have a more subtle mathematical nature \cite{Murray:1989zk}. They may change under gauge transformations, nonetheless, they describe important properties of the monopole. Holomorphic charges define a stratification of each topological sector in connected subspaces. Their value changes discontinuously from one stratum to another.


\subsection{Framed moduli spaces of monopoles}

 It is customary in literature to consider two different definitions of moduli spaces. Let us consider the set of field configurations $(A_{i},\Phi)$ which satisfy the BPS equations and the boundary conditions (\ref{eq:higgsboundary}) and (\ref{eq:boundary}).  The stabilizer for $\Phi_{\infty}$ is defined  as the set of ``unframed'' gauge transformations which leave the adjoint field invariant (along some arbitrary chosen $x_3$ direction):
\beq
\mathcal S_{0}(x) \in G : 
\quad\mathcal S_{0}\,  \Phi (0,0,x_{3}) =\Phi(0,0,x_{3}).
\eeq
The moduli space space of ``unframed'' monopoles is thus defined as the set of gauge inequivalent configurations:
\beq
\mathcal M_{\rm unframed}=\left\{    (A_{i},\Phi)|(\ref{eq:boundaryphi}) \right\}\,  / \, \mathcal S_{0}       .
\label{eq:unframed}
\eeq
For a fundamental $SU(2)$ monopole, for example, it is just  given by moduli related to spatial translations:
\beq
\mathcal M_{\rm unframed}^{SU(2),k=1}=\mathbb R^{3}.
\label{eq:unframedsu}
\eeq
However, it was soon realized, for the 't Hooft-Polyakov monopole, that the $S^{1}$ phase generated by the $U(1)$ stabilizer has an important physical effect. In fact, upon quantization, this phase gives rise to an infinite tower of electrically charged states (dyons) \cite{Julia:1975ff}. The moduli space of the most physical interest is thus that of ``framed'' monopoles:
\beq
\mathcal M_{\rm framed}=\left\{    (A_{i},\Phi)|(\ref{eq:boundaryphi})\right\}  \, / \, \mathcal G_{0}     ,   
\label{eq:framed}
\eeq
where the quotient is only  taken with framed gauge transformation $ \mathcal G_{0}$
\beq
\mathcal G_{0}(x) \in G: \quad \mathcal G_{0}(0,0,x_{3}) \rightarrow 1, \quad x_{3}\rightarrow\infty.
\eeq
In the $SU(2)$ case, this definition correctly includes the relevant $S^{1}$ phase:
\beq
\mathcal M_{\rm framed}^{SU(2),k=1}=\mathbb R^{3}\times S^{1}.
\eeq

The framed moduli spaces of  monopoles capture important new features which arise in the non-Abelian case. Unframed gauge transformations $\mathcal S_{0}$ generically do not leave the quantity $G_{0}$ invariant, and generate what may be called a ``magnetic orbit'' \cite{Bais:1997qy}. The existence of these orbits gives rise to various subtleties in the quantization of non-Abelian monopoles. As can be intuitively seen from Eq. (\ref{eq:higgsboundary}) the modes generated by variations of $G_{0}$ are non-normalizable, behaving as $\sim 1/r$ \cite{Abouelsaood:1982dz,Abouelsaood:1983gw}. This is problematic when one tries to apply standard quantization methods to the non-Abelian modes. Another related problem is the non-existence of  a stabilizer  $\mathcal S_{0}$ which is globally defined on the whole two-sphere at spatial infinity \cite{Nelson:1983bu,Balachandran:1982gt}. Moreover, as noticed in \cite{Bais:1997qy,Murray:1989zk}, the physical interpretation of holomorphic charges is also not completely clear. A reason for this is that monopole configurations with certain holomorphic charges cannot be considered as composite state of fundamental objects.

It is widely believed that a correct physical understanding of the moduli space of non-Abelian monopoles and its quantization  would shed more light into the issue of dualities in non-Abelian gauge theories.
We are not concerned here in this important issue, but rather we limit ourselves to consider the structure of the classical moduli space.


\subsection{Rational map construction for the monopole moduli space}

The rational map construction is based on the existence of a one-to-one correspondence  between the moduli space of framed monopoles and the space of based rational maps from $\mathbb C P^{1}$ to a special class of flag manifolds\footnote{With based we mean that a chosen point of $\mathbb CP^{1}$ is mapped into a chosen point of the flag, for every map.} \cite{Murray:1989zk,Jarvis,Jarvis2}:
\beq
R(z): \quad \mathbb CP^{1} \rightarrow \mathbb Flag_{n_{1},\dots,n_{q}}\,.
\label{eq:ratdef}
\eeq
These flag manifold are given by the following homogeneous spaces
\beq
\mathbb Flag_{n_{1},\dots,n_{q}}=\frac{SU(N)}{S(U(n_{1})\times\dots \times U(n_{q}))}\,.
\label{eq:flagdef}
\eeq
The quotient above can also be expressed in a complexified form:
\beq
\mathbb Flag_{n_{1},\dots,n_{q}}=\frac{SL(N,\mathbb C)}{P_{n_{1},\dots,n_{q}}},
\eeq
where $P_{n_{1},\dots,n_{q}}$ is the parabolic group given by the set of upper-block-triangular complex matrices
\[
P_{n_{1},\cdots,n_{q}}=\left(
\begin{array}{c|c|c}
  P_{n_{1}}& \cdots  &  \cdots \\
  \hline
 0 &  \ddots &   \vdots \\
  \hline
 0 & 0  &   P_{n_{q}}
\end{array}
\right).
\label{eq:parabolic}
\]
The quotient above can be realized by right side multiplication of an invertible matrix $f \sim f P$, and we can completely fix $P$ by putting $f$ into a lower-block triangular matrix, which gives a set of coordinates for the flag manifold
\[
f^{l}=\left(
\begin{array}{c|c|c}
  {\bf 1}_{n_{1}}& 0 &  0 \\
  \hline
 F_{1} &  \ddots &   0 \\
  \hline
 F_{2} & F_{3}  &   {\bf 1}_{n_{q}}
\end{array}
\right).
\]
A based holomorphic rational map $R(z)$ is obtained by promoting the elements of $f^{l}$ to be ratios of polynomials. The condition that the map is based can be realized by imposing, for example, that all the ratios go to zero at large values of $z$
\beq
R(z): \quad f^{l}(z)=\left(
\begin{array}{c|c|c}
  {\bf 1}_{n_{1}}& 0 &  0 \\
  \hline
 R_{1}(z) &  \ddots &   0 \\
  \hline
 R_{2}(z) & R_{3}(z)  &   {\bf 1}_{n_{q}}
\end{array}
\right), \quad R_{i}(z) \rightarrow 0, \quad z \rightarrow \infty.
\eeq
A crucial property of these rational maps is that their moduli space is partitioned in terms of topological numbers and stratified in terms of holomorphic integers exactly as the monopole moduli space \cite{Murray:1989zk,BHMM}.

Let us now be more concrete by considering the $SU(2)$ case originally considered by Donaldson \cite{Donaldson:1985id}. We will follow Ref. \cite{Manton:2004tk} to describe the more constructive approach of Hurtubise \cite{Hurtubise:1985vq}, which makes use of the scattering data of the Hitchin equation \cite{Hitchin:1982gh}. Let us first consider an oriented straight line. This line will be parametrized by the coordinate  $x_{3}$, while the orthogonal plane is described by the complex coordinate $z$. For each point on the plane, then, we consider the following Hithcin equation:
\begin{equation}
\nabla_{H}\psi=(D_{3}+\Phi)\psi=0
\label{eq:hitchin}
\end{equation}
where $\psi$ is a 2-component {complex} spinor field. The asymptotic value of $\Phi$ can be chosen to be proportional to $\tau^{3}$:
\beq
\Phi_{\infty}=2 \mu \, \tau_{3}= 
\left(
\begin{array}{cc}
  \mu&   0\\
  0  & -\mu   
\end{array}
\right)\,.
\eeq
The Hitchin equation has 2 independent solutions, whose exponential behavior is dictated by the eigenvalues of $\Phi$. We thus have solutions which decay and grow exponentially for large $|x_{3}|$. We can approximately solve the Hitchin equation at large values of  $|x_{3}|$ using the asymptotic expression given by Eq. (\ref{eq:higgsboundary}):
\begin{equation}
\Phi= \left(2 \mu-\frac{m}{ g\, |x_{3}|}    \right)\tau^{3}+\mathcal O\left(\frac{1}{|x_{3}|^{2}}\right), 
\end{equation}
where $m$ is the monopole number.
The solution which decay at large positive $x_{3}$ is thus asymptotically given by:
\begin{equation}
\psi(x_{3})\sim  \left(
\begin{array}{c}
   0  \\
   1
\end{array}
\right)|x_{3}|^{m/g}e^{- \mu \, x_{3}}.
\end{equation}
The same solution can be generally expressed at large negative $|x_{3}|$ in terms of the scattering coefficients $a$ and $b$:
\begin{equation}
\psi(x_{3})\sim b \left(
\begin{array}{c}
    1  \\
    0
\end{array}
\right)|x_{3}|^{m/g}e^{ \mu \, x_{3}}+a \left(
\begin{array}{c}
   0  \\
   1
\end{array}
\right)|x_{3}|^{m/g}e^{- \mu \, x_{3}}.
\label{eq:scatt}
\end{equation}
We are interested in the scattering coefficients as functions of $z$. First, we observe that the Hitchin operator commutes with the complex covariant derivative $D_{\bar z}$
\beq
D_{\bar z}\equiv\frac12 (D_{1}+i\,D_{2}).
\eeq
 This follows from the BPS equations (\ref{eq:bogom})
\begin{equation}
[D_{\bar z}, \nabla_{H}]=-i(F_{13}+i F_{23})+(D_{1}+i D_{2})\Phi=0\,.
 \label{eq:integrability}
\end{equation}
At large values of  $|x_{3}|$, in the gauge choice made in Eq. (\ref{eq:higgsboundary}), the covariant derivatives reduce to ordinary ones, $D_{\bar z}\rightarrow \partial_{\bar z}$, and from the equations above it follows that
\begin{eqnarray}
& \partial_{\bar z }\nabla_{H}\psi=\nabla_{H}\partial_{\bar z}\psi=0,&  \nonumber \\
 & \Downarrow & \nonumber \\
 &  \partial_{\bar z}\psi=0,& 
\end{eqnarray}
which implies that the scattering coefficients $a$ and $b$ are holomorphic functions of the variable $z$:
\beq
 a \rightarrow a(z),\quad  b \rightarrow b(z).
\eeq 
At large distances from the monopole, $ |z|\rightarrow \infty$, the Hitchin equation is trivial. This implies  the boundary values for the coefficients:
\begin{equation}
b(z)/a(z) \rightarrow 0, \quad z \rightarrow \infty.
\label{eq:ratbound}
\end{equation}
This condition, together with the assumption of analyticity  and continuity,  implies that $a(z)/b(z)$ is a ratio of polynomials\footnote{It is possible to show that even at arbitrary small values of $|x_{3}|$ the ratio $R(z)=b(z)/a(z)$ is an holomorphic function of $z$}:
\beq
R(z) \equiv \frac{b(z)}{a(z)}=  \frac{Q(z)}{P(z)}=\frac{q_{1}z^{m-1}+\dots+q_{m}}{z^{m}+p_{1}z^{m-1}+\dots+p_{m}}.
\eeq
The degree of the polynomials is determined by the monopole number $m$, which is also the total magnetic charge of the corresponding configuration. The ratio $R(z)$ is independent of the normalization of the solutions. It contains $2 m$ complex variables, which define the monopole moduli space. The expression above realizes the rational map construction for $SU(2)$ monopoles, which according to Eq. (\ref{eq:ratdef}) in this case is a map from $\mathbb CP^{1}$ to  $\mathbb CP^{1}$. 

The Donaldson construction gives a simple prescription for the determination of the monopole moduli space. It is quite non-trivial, however, to reconstruct the explicit solution starting by the data of the rational map. To this end, it is still more convenient to use the Nahm transform. The choice of a particular direction also hides most of the symmetries  that a  monopole configuration may have. For example, rotational symmetry is not manifest. The action of general symmetry transformations on the Donaldson rational map is still not known explicitly. There is an exception for translations and symmetries not broken by the choice of a line: rotations along the $x_{3}$-axis and reflections. Rotations of an angle $\theta$ act on the rational map in the following way
\beq
\gamma : R(z) \quad\longrightarrow\quad R(e^{i \theta} z),
\eeq
while  translations are given by
\beq
\delta : R(z) \quad \longrightarrow \quad e^{-\delta x_{3} \,m/g }\, R(z-\delta z)\,.
\eeq 
Reflections $\sigma : (z,x_{3}) \rightarrow (z,-x_{3})$ act in a more involved way:
\beq
\sigma : R=\frac{Q}{P} \quad \longrightarrow \quad \tilde R=\frac{\tilde Q}{P},
\eeq
where $\tilde Q$ is given by the following relation: 
\beq
Q \tilde Q=1.
\eeq
The monopole configuration is invariant if and only if the action of one of the above symmetry transformations is equivalent to a change of the framing of the map \cite{Manton:2004tk}:
\beq
R_{i}(z) \quad \longrightarrow\quad e^{i \alpha} R_{i}(z).
\eeq

The generalization to the non-Abelian case with minimal symmetry breaking is straightforward. The spinor $\psi$ will now have $N$ components, as many as the independent solutions of the Hitchin equation. The rational map will be given by a collection of $N-1$ ratios $R_{i}(z)\equiv b_{i}(z)/a(z) \,(i=1,\cdots,N-1)$. The action of the unframed gauge transformations $ \mathcal S_{0}$ on the scattering coefficients can also be easily determined:
\beq
 b_{i} \rightarrow U_{ij}b_{j}, \quad U \in \mathcal S_{0}=SU(N-1)\times U(1).
\eeq  

Let us consider, for example, the rational map for a single $SU(N)$ monopole:
\beq
R^{1}_{i}=\frac{b_{i}}{z-z_{0}}.
\label{eq:ratmono}
\eeq
By the action of the transformations described above, it is possible to reduce the above map to the following form:
\beq
R^{1}_{1}=\frac1z, \quad  R^{1}_{i}=0, \quad i\neq 1.
\eeq 
which describe an embedded, unframed,  $SU(2)$ 't Hooft-Polyakov monopole sitting at the origin. By using this information, it is possible to determine the physical meaning of the scattering coefficients $b_{i}$. The position of the monopole is given by:
\beq
(z_{0}, x_{3,0})=\left(z_{0},\frac1{2gN}\ln(|b_{1}|^2+\dots+|b_{N-1}|^2)\right),
\eeq
while the set of ratios $b_{i}/b_{j}$ ($N-2$ of them are independent) parameterize an $S^{1}$ fibration of $\mathbb CP^{N-2}$. We have the following result for the moduli space of fundamental $SU(N)$ monopole in the minimal symmetry breaking case:
\beq
\mathcal M^{SU(N),k=1}_{\rm framed} =\mathbb R^{3}\times (S_{\rm el}^{1}\ltimes \mathbb CP_{\rm mag}^{N-2})= \mathbb R^{3}\times S^{2N-3}.
\label{eq:modspaceminimal}
\eeq
In the expression above we have used the terminology of Ref. \cite{Bais:1997qy} to distinguish between an ``electrical'' orbit generated by the Abelian factor and a  ``magnetic'' orbit generated by the non-Abelian residual gauge symmetry. It is important to notice that the electric orbit is non-trivially fibered on the magnetic one. This fact is related to the difficulties to associate to non-Abelian monopoles well-defined algebraic objects, as required by quantum mechanics \cite{Bais:1997qy,Kampmeijer:2008wz,Kampmeijer:2008hw}.

\subsection{Composite monopoles in $SU(3)$ gauge theory:}
\label{sec:compmon}

In this section we review in detail the moduli space of two $SU(3)$ monopoles, by  explicitly constructing the associated rational maps. We follow the approach of \cite{BHMM}.

\subsubsection*{Maximal Breaking: $SU(3)\rightarrow U(1)\times U(1)$ }
The moduli space of monopoles in this case is related to based rational maps into the following flag manifold:
\beq
\mathbb F lag_{1,1}=\frac{SU(3)}{U(1)\times U(1)}\,.
\eeq
As well known from mathematical literature, this space can be obtained in terms of a  complexified  quotient
\beq
\mathbb F lag_{1,1}=SL(3,\mathbb C)/B_{1,1}\,,
\eeq
where $B_{1,1}$ is the Borel group of upper triangular matrices. The space  $\mathbb F lag_{1,1}$ can thus be realized as the set of 3 by 3 invertible matrices $M$, quotiented by the right action of $B_{1,1}$:
\beq
\mathbb F lag_{1,1}=\{ M, \,\, M\sim MB_{1,1};\,\, M \,\, {\rm invertible}\}.
\eeq
Then we can completely fix $B$ by putting the  matrix $M$  into the following lower triangular form:
\beq
M=\left(
\begin{array}{ccc}
  1& 0  & 0  \\
  a&  1 & 0  \\
   c& b  &  1 
\end{array}
\right).
\label{eq:M}
\eeq
A based rational map can be simply obtained by promoting the coefficients $a,b$ and $c$ to be holomorphic functions of $z$  which vanish at infinity. The study of a composite configuration of two monopoles with the same charges, $(m_{1},m_{2})=(2,0)$ or $(m_{1},m_{2})=(0,2)$, can be reduced to that of $SU(2)$ monopoles. Because of this,  here we just consider a composite monopole configuration with the following magnetic charge $(m_{1},m_{2})=(1,1)$. 

Since the rational map construction requires the choice of a preferred spatial direction we have two possibilities: the two monopoles can be aligned along the $x_{3}$ or separated on the $z$ plane. Let us start with the case of non-aligned monopoles. The correct rational map is given, as shown in Ref. \cite{BHMM}, by 
\beq
R_{(1,1)}^{\rm sep}(z)=
\left(
\begin{array}{ccc}
 1 & 0  & 0  \\
 \beta/(z-z_{1}) & 1  & 0  \\
 0 &     \alpha/(z-z_{2}) &   1
\end{array}
\right),\quad \alpha,\beta\in \mathbb C^{*} .
\label{eq:compmonopsep}
\eeq
It represents a $(0,1)$ monopole located at the point $z_{1}$ and a $(1,0)$ monopole located at $z_{2}$. There is no $(1,0)$ or $(0,1)$ monopole if either $\alpha$ or $\beta$ is zero. The moduli space is then:
\beq
\mathcal M^{\rm mon}_{(1,1){\rm sep}}=\left( \mathbb C \times\mathbb C^{*} \right)^{2}\,.
\eeq
The rational map representing a configuration of two aligned  $(1,1)$ monopoles is, instead:
\beq
R_{(1,1)}^{\rm align}(z)=
\left(
\begin{array}{ccc}
 1 & 0  & 0  \\
 \beta/(z-z_{0}) & 1  & 0  \\
 \gamma/(z-z_{0}) &     \alpha/(z-z_{0}) &   1
\end{array}
\right),\quad \alpha,\beta\in \mathbb C,\,\gamma \in \mathbb C^{*}, \quad  \alpha\,\beta=0.
\label{eq:compmonop}
\eeq
The conditions on the parameters are needed to ensure that the map describes the  right topological sector. Foe example,  if $\alpha\,\beta\neq0$, the topological number will be $(2,1)$. On the other hand, setting $\gamma$ to zero would reduce the topological numbers to $(0,1)$ or $(1,0)$, in the case  $\alpha$ or $\beta$ vanishes respectively. The two cases with either $\alpha$ or $\beta$ equal to zero can be interpreted as  representing monopoles aligned along $x_{3}$ with a different order. Setting $\alpha=\beta=0$ gives in fact a  configuration of coincident monopoles. The moduli space in this case has complex dimension 3 and it is given by:
\beq
\mathcal M^{\rm mon}_{(1,1){\rm align}}=\left\{ \mathbb C(z_{0})\times\mathbb C^{2}(\alpha,\beta)\times \mathbb C^{*}(\gamma)\; | \;\alpha\beta=0 \right\}
\eeq

\subsubsection*{Minimal Breaking: $SU(3)\rightarrow SU(2)\times U(1)$ }
When two eigenvalues of $\Phi$ coincide, say $\mu_{1}=\mu_{2}$, the unbroken symmetry is enhanced, and become non-Abelian. Moreover, of the two magnetic charges $(m_{1},m_{2})$, only $m_{2}$ remains topological, while $m_{1}$ is holomorphic. As a consequence, all the examples of the previous section degenerate into a fundamental non-Abelian  monopole $(0,1)$.

 Let us describe in detail what happens to the monopoles moduli space as we start from the maximally broken case of the previous section and we tune two masses to be coincident. The relevant flag is now
\beq
\mathbb F lag_{2,1}=SL(3,\mathbb C)/B_{2,1}\,,
\eeq
where $B_{2,1}$ is the upper block-triangular:
\[
B_{2,1}=
\left(
\begin{array}{c|cc}
 i & j & k  \\
 \hline
  0 & l  & m  \\
  0 & n  & o  
\end{array}.
\right)
\]
It is easy to see that, with this enhanced symmetry we can always put $b=0$ in \ref{eq:M}. This implies  $\alpha=0$ in Eqs. (\ref{eq:compmonopsep}) and (\ref{eq:compmonop}). We can obtain the most general configuration for a fundamental non-Abelian monopole from the aligned case Eq. (\ref{eq:compmonop}) only:
\beq
R_{1}(z)=
\left(
\begin{array}{ccc}
 1 & 0  & 0  \\
 \beta/(z-z_{0}) & 1  & 0  \\
 \gamma/(z-z_{0}) &     0 &   1
\end{array}
\right),\quad \beta,\gamma \in \mathbb C^{2}\backslash\{0,0\}.
\label{eq:compmonop2}
\eeq
Notice that is now allowed to have $\gamma=0$\footnote{The monopole corresponding to (\ref{eq:compmonopsep}) will reduce to a special non-Abelian monopole with $\gamma=0$. }. Setting $\gamma$ to zero changes the value of the holomorphic charge $m_{1}$ from 1 to 0, but leaves the topological charge unmodified. In physical terms, when $\gamma=0$ the $(1,0)$ monopole in the maximally broken case is sent to spatial infinity on the $x_{3}$ line. The moduli space can then be written as:\footnote{The moduli space in the non-Abelian case can be considered as a patching of the moduli spaces of the Abelian monopoles  $(1,1)$ and $(0,1)$ described in the previous section.}:
\beq
\mathcal M^{\rm mon}_{(1)}=\left\{ \mathbb C(z_{0})\times\mathbb C^{2}(\beta,\gamma)\backslash\{0,0\} \right\}\sim  \mathbb R^{3} \times S^{3}\,,
\eeq
which is the same result written in  Eq.~(\ref{eq:modspaceminimal}).\footnote{The parameters $\alpha$ and $\beta$ correspond to the scattering coefficients $b_{i}$ in the previous construction of the rational map.}


\section{Non-Abelian Monopoles in the Higgs Phase}



 The most natural and convenient way to  embed the monopole theory (\ref{eq:Mon act}), in the BPS saturated case, is to consider  a $U(N)$ gauge theory with extended $\mathcal N=2$ supersymmetry. To ensure the existence of a supersymmetric vacuum, we also include $N_{F}=N$ fundamental hypermultiplets. The bosonic part of this model is\footnote{The complete bosonic sector includes further $N$ anti-fundamental multiplets $\tilde q^{A}$. We set them to zero, $\tilde q^{A} = 0$, as they are trivial in  the classical configurations we are going to discuss.}:
\begin{eqnarray}
S & = & \int d^{4}x \left\{   \frac1{4 g^{2}}(F^{0}_{\mu\nu})^{2}+\frac1{4 g^{2}}(F^{a}_{\mu\nu})^{2}+\frac{1}{g^{2}} |\partial_{\mu}\phi^{0}|^{2}+\frac{1}{g^{2}} |D_{\mu}\phi^{a}|^{2}+ |\nabla_{\mu} q^{A}|^{2} +\right. \nonumber \\
 & + &  \frac{g^{2}}{2}\left(  \frac{1}{g^{2}}\epsilon^{abc}\bar \phi^{b} \phi^{c}+\bar q^{A}\frac{\tau^{a}}{2}q^{A}  \right)^{2 }+\frac{g^{2}}8(|q^{A}|^{2}-N\xi)^{2} +\nonumber \\
 & + & \left.\frac12\left| (\phi^{0}\frac{2}{\sqrt{2N}}+\phi^{a}\tau^{a}+\sqrt2 m_{A})q^{A}  \right|\right\}\,,
 \label{eq:kinkact}
\end{eqnarray}
with:
\begin{equation}
A=1,2, \dots, N_{F}\quad \quad \nabla_{\mu}=\partial_{\mu}-\frac{i}{\sqrt{2N}}  A^{0}_{\mu}-i \frac{\tau^{a}}2 A^{a}_{\mu}\,.
\end{equation}
The real parameter $\xi$ is the Fayet-Iliopolous term \cite{Fayet:1974jb}. As we will see shortly, a non-zero $\xi$ puts the theory into the Higgs phase. 
If  the masses $m_{A}$ are taken real, we  can consistently consider the adjoint fields $a^{0}, \, a^{a}$ to be real as well on the  solitonic solutions. The above expression then simplifies:
\begin{eqnarray}
S & = & \int d^{4}x \left\{   \frac1{4 g^{2}}(F^{0}_{\mu\nu})^{2}+\frac1{4 g^{2}}(F^{a}_{\mu\nu})^{2}+\frac{1}{g^{2}} |\partial_{\mu}\phi^{0}|^{2}+\frac{1}{g^{2}} |D_{\mu}\phi^{a}|^{2}+ |\nabla_{\mu} q^{A}|^{2} +\right. \nonumber \\
 & + &  \left. \frac{g^{2}}8\left(\bar q^{A}\tau^{a}q^{A}  \right)^{2 }+\frac{g^{2}}8(\bar q^{A}q^{A}-N\xi)^{2} +\ \frac12\left| (\phi^{0}\frac{2}{\sqrt{2N}}+\phi^{a}\tau^{a}+\sqrt2 m_{A})q^{A}  \right| \right\}.
 \label{eq:kinkactred}
\end{eqnarray}
We can fit all the fields into $N\times N$ matrices
\begin{equation}
F_{\mu\nu}\equiv F^{0}_{\mu\nu} \frac{{\bf 1}_N}{\sqrt{2N}} +F^{a}_{\mu\nu}\frac{\tau^{a}}{2}, \quad \Phi\equiv\sqrt2\left(\phi^{0}\frac{{\bf 1}_N}{\sqrt{2N}}+\phi^{a}\frac{\tau^{a}}{2}\right), \quad Q\equiv q_{i}^{A},
\end{equation}
 in terms of which the action (\ref{eq:kinkactred}) can be written in the following compact form
\beq
S  =  \int d^{4}x \, \Tr\left\{   \frac1{2 g^{2}}F_{\mu\nu}^{2}+\frac{1}{g^{2}} |D_{\mu}\Phi|^{2}+ |\nabla_{\mu} Q|^{2} +\right.   \left.\frac{g^{2}}4(Q \bar Q -\xi)^{2} +\left|\Phi Q+ Q  M  \right|^{2} \right\}\,,
\label{eq:kinkreduced}
\eeq
where we have defined the mass matrix $M$ as:
\beq
M_{AB}=\delta_{AB}m_{A}=
\left(
\begin{array}{cccc}
  m_{1} & 0 & \cdots  & 0  \\
  0 & m_{2}  & \cdots & 0  \\
  \vdots & \cdots & \ddots & \vdots \\
  0 & \cdots  & \cdots & m_{N}  
\end{array}
\right), \quad m_{1}\le m_{2}\le\dots\le m_{N}.
\eeq
Non-zero masses generically break the $SU(N)_{\rm F}$ flavor symmetry down to $U(1)_{\rm F}^{N-1}$.  Notice that we can always absorb an equal contribution to the masses into a shift of the adjoint field $\Phi$. With no loss of generality thus, we can always set $\sum_{A=1}^{N}m_{A}=0$.

This model has a unique vacuum (up to gauge-flavor symmetry transformations)
\beq
\Phi_{0}=-M, \quad Q=\sqrt\xi \,  1_{N}
\eeq
 invariant under a  ``color-flavor locked'' global  symmetry $H_{\rm C+F}$ which plays an important role in the study of the moduli space of solitons:
\beq
H_{\rm C+F}(\Phi)=H_{\rm C}\Phi H^{-1}_{\rm C}= \Phi  \quad H_{\rm C+F}(Q) \equiv H_{\rm C}\,Q\,H_{\rm F}^{-1}, \quad H_{\rm C}= H_{\rm F}\,.
\eeq
This residual color-flavor symmetry is determined by the vacuum value of $\Phi$. In the most general case, it is given by the stabilizer of the adjoint field, similarly to what happens in the unbroken phase (see Eq. (\ref{eq:nonmaximal})):
\beq
H_{\rm C+F}=S(U(n_{1})\times U(n_{2})\dots\times U(n_{q})),
\label{eq:colflav}
\eeq
in the case where there are $q$ {sets} of fields with degenerates masses.
 The theory has two parameters with non-trivial mass dimension $m\sim m_{i}$ and $\sqrt\xi$.\footnote{For convenience we chose the masses $m_{i}$ to be all of the same order $m$.} Playing with the relative value of these two parameters we can put the theory in  two different regimes. When  $m\gg \sqrt\xi$, the symmetry breaking in the vacuum is:
\beq
U(N)_{\rm C}\times SU(N)_{\rm F}\stackrel{m}{\longrightarrow}U(1)_{\rm C}\times H_{\rm C}\times H_{\rm F}\stackrel{\sqrt\xi}{\longrightarrow} H_{\rm C+F}.
\label{eq:coul}
\eeq
It supports ``almost free'' monopoles, with a typical size of order $\Delta m^{-1}$, confined by very wide flux tubes with a width of order $\sqrt\xi^{-1}$. On the other hand, when $m\ll \sqrt\xi$ we have:
\beq
U(N)_{C}\times SU(N)_{F}\stackrel{\sqrt\xi}{\longrightarrow}SU(N)_{C+F}\stackrel{m}{\longrightarrow} H_{C+F}.
\label{eq:Higgs}
\eeq
In this second regime, the theory support $N$, very narrow, flux tubes. Monopoles are now squeezed into the flux tubes, and correspond to kinks, interpolating between different various string-like solitons \cite{Hanany:2004ea,Shifman:2004dr}.
 When two or more masses are degenerate,  the regime (\ref{eq:coul}) will supports confined non-Abelian monopoles, while the regime (\ref{eq:Higgs}) will supports $q$ different kinds of  1/2 BPS non-Abelian vortices and 1/4 kinks interpolating among them\footnote{The theory admits a whole zoo of solitons, including domain walls, junctions of vortices on walls...\cite{Shifman:2007ce,Eto:2006pg}}. 
 
It is quite remarkable that the solitons in the two regimes can be studied by analyzing the same set of Bogomol'nyi equations, which follow from the square root completion the action (\ref{eq:kinkreduced}) \cite{Tong:2003pz,Hanany:2004ea,Shifman:2004dr,Isozumi:2004vg,Eto:2006pg}:
\begin{eqnarray}
S & = & \int d^{3}x\, \Tr\left\{  \frac1{g^{2}}\left( F_{12}-D_{3}\Phi + \frac{g^{2}}2(Q \bar Q -\xi)\right)^{2} +\right. \nonumber \\
 & + & |\nabla_{1} Q+i \nabla_{2} Q|^{2}  +  \nonumber \\ 
 & + &\frac1{g^{2}}\left(D_{1}\Phi-F_{23}\right)^{2}+\frac1{g^{2}}\left( D_{2}\Phi+F_{13}\right)^{2}  \nonumber \\ 
 & + & \left| \nabla_{3}Q +  \Phi Q+  Q  M  \right|^{2} \nonumber \\
 & + & \left. \xi\,F_{12}+\frac1{g^{2}}\partial_{i}(\epsilon_{ijk}\Phi \,F_{jk}) \right\}.
 \label{eq:boghiggs}
\end{eqnarray}
The Bogomol'nyi equations are given, as usual, by imposing vanishing of the positive definite contributions:
\begin{eqnarray}
 \nabla_{1} Q+i \nabla_{2} Q & = & 0  \nonumber \\ 
 \nabla_{3}Q +  \Phi Q+  Q  M  & =& 0  \nonumber \\
 D_{1}\Phi-F_{23}=0, \quad  D_{2}\Phi+F_{13} & = & 0 \nonumber  \\
 F_{12}+ D_{3}\Phi + \frac{g^{2}}2(Q \bar Q -\xi) & = & 0
 \label{eq:kinkbps}
 \end{eqnarray}
 The total mass  is  the sum of the last two terms in the action (\ref{eq:boghiggs}):
 \begin{eqnarray}
 M & = &\int d^{3}x\, \Tr\left\{ \xi\,F_{12}+\frac1{g^{2}}\partial_{i}(\epsilon_{ijk}\Phi \,F_{jk})\right\}\,.
 \end{eqnarray}
 The first term is infinite and related to the tension of the string. It is proportional to the total Abelian flux flowing through the space:
\beq
M_{\rm vort}=  \xi \int d^{3}x\, \Tr\, F_{12} = 2 \pi \xi L \, n^{\rm (v)},
\eeq
where $n^{\rm (v)}$ is the total number of vortices and we have regulated the integration with a finite length $L$ of the vortex. The second term gives the mass of the kinks and must be evaluated as a difference  between  the non-Abelian fluxes flowing through  the   two planes at positive and negative infinity
 \begin{eqnarray}
 M_{\rm kink} & =&  \frac1{g^{2}} \int d^{3}x\, \Tr \, \partial_{i}(\epsilon_{ijk}\Phi \,F_{jk})=\frac1{g^{2}} \int d^{2}x\, \Tr ( (\Delta  B_{3})\Phi ),\nonumber \\
  \Delta B_{3}& =& B_{3}(x_{1},x_{2},x_3=+\infty)-B_{3}(x_{1},x_{2},x_3=-\infty).
 \end{eqnarray}
 The flux of  $B_{3}$ can be determined, for large $|x_{3}|$,  in terms of the vortex number. In the general case of symmetry breaking given by Eqs.~(\ref{eq:colflav}) and (\ref{eq:coul}) we have $q$ distinct topological numbers
 \beq
 n^{\rm (v)}\equiv\sum_{t=1}^{q}n_{t}^{\rm (v)}, \quad n_{t}^{\rm (v)}\equiv \frac{1}{2\pi}\int d^{2}x \, \Tr( B_{3} \, \tau_{t}^{0}),
 \eeq 
where  $\tau_{t}^{0}$ is a $U(1)$ generator  in the unbroken $t$-th sector:
 \beq
  \quad \tau_{t}^{0}\equiv
\left(
\begin{array}{c|c|c}
  0 & 0   & 0  \\
  \hline
 0  & 1_{n_{t}}  & 0  \\
  \hline
 0  &  0 &   0
\end{array}
\right)\,.
 \eeq
 The total contribution to the mass coming from kinks is given in terms of the vortex numbers 
 \begin{eqnarray}
 M_{\rm kink} & =& \frac1{g^{2}} \int d^{2}x\, \Tr ( (\Delta  B_{3})\Phi )= \frac{2 \pi}{g^{2}} \sum_t \Delta n_{t}^{\rm (v)}\phi_{t}^{0},\quad \Phi\equiv\phi_{t}^{0}\tau_{t}^{0}\,,
 \end{eqnarray}
 where the symbol $\Delta$ always indicates the difference between the values of a quantity evaluated at positive and negative infinities.
 By comparing Eq. (\ref{eq:monopolemassfinal}) with Eq. (\ref{eq:phimu}) we find a relationship between the vortex number and the monopole/kink charges:
\beq
\sum_t \Delta n_{t}^{\rm (v)}\phi_{t}^{0}=\sum_tn_{t}^{\rm (k)}\,\vec h \cdot \vec \beta_{t}^{*}= \sum_tm_{t}^{\rm top}\,\vec h \cdot \vec \beta_{t}^{*}.
\label{eq:vortexkinkrel}
\eeq
To write the relations above we have used the fact that the masses of monopole and kinks coincide \cite{Tong:2003pz,Shifman:2004dr}. Furthermore, we identify the kink numbers $n_{t}^{\rm (k)}$ with the magnetic charges of the monopoles $m_{t}^{\rm top}$.
 The number of kinks can also be easily defined in a more direct way if we recall that each fundamental kink interpolate between two vortices built in two neighboring $U(n_{t})$ blocks. We thus may write the following relations:
\beq
n_{t}^{\rm (k)}-n_{t-1}^{\rm (k)}= \Delta n_{t}^{\rm (v)}, \quad t=1,\dots,q-1, \quad n_{0}^{\rm (k)}=n_{q}^{\rm (k)}=0,
\eeq
from which we get:
\beq
n_{t}^{\rm (k)}& =& \sum_{l=1}^{t}\Delta n_{l}^{\rm (v)}\,;  \nonumber \\
n^{\rm (k)}& =& \sum_{t=1}^{q-1} n_{t}^{\rm (k)}= \sum_{l=1}^{q-1}(q-l)\Delta n_{l}^{\rm (v)}= \sum_{l=1}^{q-1}l\, \Delta n_{l}^{\rm (v)}\,.
\label{eq:kinknumber}
\eeq
The equations above can be derived, of course, by using explicit expressions for the co-roots $\vec \beta_{t}^{*}$ in Eq. (\ref{eq:vortexkinkrel}).  The relation between the magnetic flux of non-Abelian vortices and the GNOW quantization condition was first noticed in \cite{Eto:2009bg}. It can be viewed as a consequence of the fact that the same magnetic flux of the monopole must be confined by vortices \cite{Auzzi:2003em}. Here we related directly the flux of non-Abelian vortices with the magnetic charges of monopoles.


\section{Moduli Matrix Formalism for Kinks}
The BPS equations (\ref{eq:kinkbps}) were first considered in \cite{Tong:2003pz,Shifman:2004dr}. We follow the approach of Refs. \cite{Isozumi:2004vg,Eto:2005yh,Eto:2006uw,Eto:2006pg}, where  the moduli matrix technology was thoroughly developed to construct BPS solitonic configurations in the Higgs phase. 
The moduli matrix was first applied to domain walls \cite{Isozumi:2004jc,Isozumi:2004va,Eto:2004vy,Eto:2005wf,Eto:2008dm} and then extended to 
non-Abelian vortices \cite{Eto:2009bg,Eto:2005yh,Eto:2006cx,Eto:2006db,Eto:2006dx,Eto:2007yv,Eto:2008yi,Eto:2009wq,Eto:2010aj,Fujimori:2010fk} 
and BPS composite solitons \cite{Isozumi:2004vg,Eto:2004rz,Eto:2005cp,Eto:2005fm,Eto:2005sw,Eto:2006bb,Eto:2007uc}.
The moduli matrix is believed to exhaust all possible solution of the BPS equations, provided that a likely generalization of the so-called Hitchin-Kobayashi correspondence  \cite{MundetiRiera:1999fd,Baptista:2004rk} holds in in the non-compact case.

 Let us start from the equation in the first line of the system (\ref{eq:kinkbps}). It can be solved by the following ansatz \cite{Isozumi:2004vg}:
\begin{eqnarray} 
A_{\bar z}= i S^{-1}\partial_{\bar z}S, & &  Q(z,\bar z,x_{3})=S^{-1}(z,\bar z,x_{3}) H_{0}(z) P(x_{3}),
\label{eq:fieldred}
\end{eqnarray}
with $S$ and $P$ being invertible matrices depending only on the specified variables. $S$ and $H_{0}$ are defined modulo an important holomorphic ``$V$-equivalence'':
\begin{equation}
S(z,\bar z,x_{3}) \, \rightarrow \, V(z)S(z,\bar z,x_{3}), \quad H_{0}(z) \, \rightarrow \, V(z) H_{0}(z)\,,
\end{equation}
where $V(z)$ is an holomorphic matrix with determinant equal to one.
The equation in the second line is also identically satisfied, with $P$ explicitly determined in terms of $x_{3}$ and the remaining adjoint fields expressed again in terms of the matrix $S$:
\begin{eqnarray}
\partial_{3}P+P M=0 &\Rightarrow&  P=e^{-M\,x_{3}} \nonumber \\
A_{3}+i \Phi &=&  i S^{-1}\partial_{3}S\,. 
\end{eqnarray}
Remarkably, the equations on the third line are now identically satisfied with no further conditions on $S$. In fact, this equation is the integrability condition for the system given by the first two lines (see Eq.~(\ref{eq:integrability})):
\begin{equation}
[\nabla_{\bar z}\,\cdot \,,(\nabla_{3}+\Phi) \cdot +\,\cdot M]=-i(F_{13}+i F_{23})\, \cdot +(D_{1}+i D_{2})\Phi\, \cdot =0.
\end{equation}
The full set of equations (\ref{eq:kinkbps}) is now reduced to the following ``master equation'' \cite{Isozumi:2004vg,Eto:2006pg}:
\begin{equation}
4 \partial_{z}\left(\Omega^{-1}\partial_{\bar z}\Omega\right)+ \partial_{3}\left(\Omega^{-1}\partial_{3}\Omega\right)+g^{2}\left(   \Omega^{-1} \Omega_{0}-\xi   \right)=0,
\end{equation}
which is nothing but the  the last line of (\ref{eq:kinkbps}) written in terms of the gauge invariant quantities:
\begin{equation}
\Omega=SS^{\dagger}, \quad \Omega_{0}=H_{0}PP^{\dagger}H_{0}^{\dagger}
= H_0 e^{-2Mx_3} H_0^\dagger.
\label{eq:fullspace}
\end{equation}
We will assume the existence and uniqueness of the solution of the master equation\footnote{Notice that the master equation can be  solved algebraically in the $g \rightarrow \infty$ limit \cite{Hindmarsh:1992yy,Isozumi:2004vg,Eto:2006pg,Eto:2007yv}(strictly speaking, to obtain regular solutions we must consider a slight generalized model, with additional flavors, $N_{F}>N$). Moreover, numerical searches can be done, which confirm the uniqueness of the solution \cite{Auzzi:2008wm,Auzzi:2007wj,Auzzi:2007iv}.}. This assumption enables us to give a precise definition of the moduli space of the solitons supported by the model. It is given by the set of holomorphic ``moduli matrices'' $H_{0}$, defined up to $V(z)$-equivalence relations:
\beq
\bigoplus_{n_{i}^{\rm (v)},n_{i}^{\rm (k)}} \mathcal M_{n_{i}^{\rm (v)},n_{i}^{\rm(k)}}=\left\{  H_{0}\, | \, H_{0} \sim V H_{0}\right\}.
\label{eq:kinkmodulus}
\eeq
The notation above means that the full moduli space is a sum of disconnected topological sectors labelled by the number of vortices  and kinks.


\subsection{Moduli spaces for kinks}
Let us first review how to explicitly construct the moduli matrix for the vortex-kink system. We will then explain how to extract from it the numbers $n_{t}^{\rm(v)},\, n_{t}^{\rm(k)}$. 

First of all, we recall that for a generic holomorphic matrix $H_{0}$, we can completely fix the $V$-equivalence by putting the matrix in the following canonical form \cite{Eto:2005yh}:
  \beq
 H_{0}=\left(
\begin{array}{cccc}
 P_{1}(z) & Q_{1,2}(z) & \dots &  Q_{1,N} (z) \\
 0  & P_{2}(z) & \dots  &   \vdots    \\ 
  \vdots & \dots& \ddots & \vdots \\
 0 & 0 & 0 &  P_{N}(z) 
\end{array}
\right), 
\label{eq:kinkmod}
 \eeq 
 where the $Q_{j,i}(z)$  are polynomials of degree less than that of the polynomials $P_{i}(z)$. To better understand the kink configuration described by the matrix above we need, first of all, to count the number vortices at both infinities. From this knowledge, as described in the previous section, we can determine the number of kinks.  To do this is convenient, as noticed in \cite{Eto:2006pg} to interpret the combination $H_{0}(z) P(x_{3})$ introduced in Eq. (\ref{eq:fieldred}) as an $x_{3}$ dependent moduli matrix for non-Abelian vortices. Using a $V$-transformation, we can actually consider the following moduli matrix, which includes the same informations about the kink moduli space as the original matrix in Eq.  (\ref{eq:kinkmod}):
\begin{eqnarray}
 H^{u}_{0}(z,x_{3})& = & P(x_{3})^{-1}H_{0}(z) P(x_{3}) =   \nonumber \\
 & = & \left(
\begin{array}{cccc}
 P^{u}_{1}(z) & Q^{u}_{1,2}(z)e^{-(m_{1}-m_{2})x_{3}} & \dots &  \vdots \\
 0  & P^{u}_{2}(z) & \dots  &   Q^{u}_{2,N} (z)e^{-(m_{2}-m_{N})x_{3}}    \\ 
  \vdots & \dots& \ddots & \vdots \\
 0 & 0 & 0 &  P^{u}_{N}(z) 
\end{array}
\right).
\label{eq:kinkmoddepu}
\end{eqnarray}
Thanks to our choice for the ordering of the masses ($m_{i} \le m_{j}$ for $i<j$), the non-diagonal elements go to zero, or to a constant at most, for $x_{3}\rightarrow-\infty$ .
%
 The number $n_{-i}^{(\rm v)}$ of vortices  at negative infinity  is related to the degree of  $P^{u}_{i}(z)$:   
  \begin{eqnarray}
  P^{u}_{i}(z)& \sim  & z^{n_{-i}^{(\rm v)}},\quad {\rm for \, large \, } z, \nonumber \\
  n_{-}^{(\rm v)} & = & \sum_{i=1}^{N}n_{-i}^{(\rm v)}.
 \end{eqnarray}
Notice that the $n_{-i}^{(\rm v)}$ all correspond to a topological integer only in the case of maximal symmetry breaking. In the general case, the topological vortex numbers defined above Eq. (\ref{eq:kinknumber}) are given by:
\beq
n_{-t}^{(\rm v)}=\sum_{i=n_{1}+\dots+n_{t-1}+1}^{n_{1}+\dots+n_{t}}n_{-i}^{(\rm v)}
\eeq

In the same way we can look at $x_{3}\rightarrow+\infty$. In this case, the non-diagonal elements in Eq.~(\ref{eq:kinkmoddepu}) diverge. We can overcome the problem by using the $V$-equivalence to put the matrix (\ref{eq:kinkmoddepu}) into a lower triangular form 
\begin{eqnarray}
 H^{l}_{0}(z,x_{3})& = & V(z)H^{u}_{0}(z,x_{3}) =   \nonumber \\
 & = & \left(
\begin{array}{cccc}
 P^{l}_{1}(z)  & 0&  \dots &  0 \\
 Q^{l}_{2,1}(z)e^{-(m_{2}-m_{1})x_{3}}  & P^{l}_{2}(z) & \dots  &   0   \\ 
  \vdots & \dots& \ddots & \vdots \\
 \dots & Q^{l}_{N,2} (z)e^{-(m_{N}-m_{2})x_{3}}  & \dots &  P^{l}_{N}(z) 
\end{array}
\right).
\label{eq:kinkmoddepd}
\end{eqnarray}
Notice that generically $P^{u}_{i}(z)\neq P^{l}_{i}(z)$, a crucial condition to have kinks. The number of vortices $n_{+i}^{(\rm v)}$ at positive infinity is  now given by the degree of $P^{l}_{i}(z)$:
\begin{eqnarray}
  P^{l}_{i}(z)& \sim  & z^{n_{+i}^{(\rm v)}},\quad {\rm for \, large \, } z, \nonumber \\
  n_{+}^{(\rm v)} & = & \sum^{N}_{i}n_{+i}^{(\rm v)}, \quad \quad n_{+t}^{(\rm v)}=\sum_{i=n_{1}+\dots+n_{t-1}+1}^{n_{1}+\dots+n_{t}}n_{+i}^{(\rm v)}\,.
 \end{eqnarray}
Of course, the total number of vortices is conserved:
\beq
n_{-}^{(\rm v)}=n_{+}^{(\rm v)}\equiv n^{(\rm v)}.
\eeq
The number of kinks will then be determined by Eq. (\ref{eq:kinknumber}). 

It is possible to rewrite the formulas  above in terms of 
$N$-component vectors $\phi_{I}(z)$, called  
orientational vectors, defined by the following condition \cite{Eto:2005yh,Eto:2006pg}:
\beq
H_{0}(z,x_{3})\phi_{I}(z,x_{3})=0\quad  \text{Mod} \quad \prod_{i}^{N}P_{i}(z),
\label{eq:orientdef}
\eeq
The vectors $\phi_{I}$ are holomorphic functions of $z$ of degree at most $n^{(\rm v)}-1$. There are precisely $n^{(\rm v)}$ linearly independent vectors, thus $I=1,\dots,n^{(\rm v)}$. As it is clear from their defining condition, the vectors $\phi_{I}$ are defined up to a complex scaling: $\phi_{I}\sim \lambda_{I} \phi_{I}$ with $\lam \in {\bf C}^*$\footnote{There is also a $GL(\mathbb C, n^{(\rm v)})$ equivalence due to the freedom of taking linear combinations of the vectors $\phi_{I}$.}. We can define the quantities $n_{-i}^{(\rm v)}$ as the number of orientational vectors which at negative infinity are equivalent to $\phi_{0,i}^{T}\equiv(0,\dots,1,0,\dots)$, where the only non-zero element is in the $i$-th position:
\beq
n_{-i}^{(\rm v)}= \text{number of} \quad \phi_{I}^{T}(z,-\infty)\sim \phi_{0,i}^{T}=(\underbrace{0,\dots,0}_{i-1},1,0,\dots).
\eeq
Similarly we define the vortex numbers at positive infinity:
\beq
n_{+i}^{(\rm v)}= \text{number of} \quad \phi_{I}^{T}(z,+\infty)\sim \phi_{0,i}^{T}=(\underbrace{0,\dots,0}_{i-1},1,0,\dots).
\eeq
Each vector can be interpreted as describing the orientation of a non-Abelian vortex. Comparing the expression of each $\phi_{I}$ at the two infinities gives us the number of kinks supported by each vortex.  If we have:
\begin{eqnarray}
\phi_{I}^{T}(z,-\infty)\sim \phi_{0,i}^{T}=(\underbrace{0,\dots,0}_{i-1},1,0,\dots), \quad \phi_{I}^{T}(z,+\infty)\sim \phi_{0,j}^{T}=(\underbrace{0,\dots,0}_{j-1},1,0,\dots), \quad i\le j, \nonumber \\
\end{eqnarray}
the number of kinks supported by the $I$-th vortex will be:
\begin{eqnarray}
n_{tI}^{\rm (k)}& =& 0, \quad  n_{1}+\dots + n_{t}<i,\nonumber \\
 n_{tI}^{\rm (k)}& =& 0, \quad j<n_{1}+\dots + n_{t+1},\nonumber \\
 n_{tI}^{\rm (k)}& =& 1, \quad {\rm otherwise}.
\label{eq:kinknumber2}
\end{eqnarray}
Again, it is easy to see that this definition gives the same result with Eq. (\ref{eq:kinknumber}).

\subsubsection*{Neutral Vortices}
The reformulation in terms of orientational vectors  enables us to give a precise definition of  what we call ``neutral vortices''.  Neutral vortices do not support any kink, and can be decoupled completely by our system without changing the number and the type of kinks. Being able to remove these vortices  is particularly useful when we compare the moduli space of kinks to the moduli space of monopoles in the unbroken phase, where vortices simply disappear from the spectrum. 
%
%

We can identify two conditions for the existence of neutral vortices, that are relevant for the purpose of this paper. The first one requires  that the  moduli matrix (\ref{eq:kinkmod}) has a column with a common factor:
\beq
H_{0}(z,x_{3})=\left(
\begin{array}{cccc}
  \ddots & \dots  & p_{i}(z)Q_{1,i}'(z)e^{-(m_{1}-m_{i})x_{3}}   &\dots \\
 \vdots &  \ddots  &  \vdots  & \vdots\\
   \vdots & \dots  & p_{i}(z)P'_{i}(z)   &  \vdots  \\
   \vdots&  \dots   &  \vdots  &   \ddots
\end{array}
\right).
\label{eq:neutralmatrix}
\eeq
Then, there exists  the following orientational vectors:
\beq
\phi_{i,m}(z)^{T}=(0,\dots,z^{m}P_{i}'(z),\dots,0), \quad 0\le m< \text{deg}\,(p_{i}).
\label{eq:neutralvec}
\eeq
The orientational vectors above do not depend on the coordinate $x_{3}$, and remain the same at both infinities. This implies that the corresponding vortex does not support any kink. In fact, this same number of vortices cancels in the differences in Eq. (\ref{eq:kinknumber}). As a general statement, in fact, a neutral vortex corresponds to an $x_{3}$-independent orientational vector. In the maximally broken case, the condition in (\ref{eq:neutralmatrix}) is sufficient to identify all neutral vortices. In the degenerate case, we have an additional situation. Consider, for example, the following moduli matrix:
\[
H^{u}_{0}(z,x_{3})=\left(
\begin{array}{c|c|c}
  H_{1}^{u}(z,x_{3}) & 0  &   \vdots \\
  \hline
  0&  H_{2}^{u}(z)   & \vdots \\
\hline
  0 & 0  & H_{3}^{u}(z,x_{3})
\end{array}
\right),
\]
where $H_{2}^{u}$ is an upper triangular moduli matrix located in a block corresponding to fields with degenerate masses. The elements of $H_{2}^{u}$ will not depend on $x_{3}$, and the same will be for the corresponding orientational vector:
\beq
\phi(z)^{T}=(\underbrace{0, \dots, 0}_{J}, \phi_{2}(z)^{T},\underbrace{0, \dots, 0}_{N-J-n_{j}}), \quad J=\sum_{i}^{j-1}n_{i}\,, 
\eeq
where $\phi_{2}(z)$ is an orientational vector for the moduli matrix $H_{2}^{u}$.
In the following, we will always simplify the moduli matrix by removing the neutral vortices described above.


\subsection{Coincident Kinks in  $U(3)$ gauge theory}

Let us now study the case of two kinks which corresponds to the monopoles of section \ref{sec:compmon}. To be able to do so, we have to consider both the cases where the two kinks are separated on the $z$ plane (confined by distinct vortices) and where two kinks are aligned along $x_{3}$ (confined by a single vortex).

\subsubsection*{Maximal breaking: $U(3)\rightarrow U(1)^{3}$}
As already explained, the number of vortices appearing in the Higgs phase is in principle arbitrary and determined by boundary conditions independent from the presence of kinks. Generically, the minimum number of vortices should be at least equal to the total number of kinks. Nevertheless,  several kinks  can be confined by a single vortex. In the Higgs phase they corresponds to a bead of at most $N-1$ kinks (see Refs. \cite{Hindmarsh:1985xc,Eto:2006pg}).

We start from a configuration of two kinks separated on the $z_{0}$ plane. According to the counting in Eq. (\ref{eq:kinknumber2}), we have to take the following moduli matrix:
\beq
H^{u}_{(1,1){\rm sep}}(z,x_{3})=
\left(
\begin{array}{ccc}
  1& \alpha\, e^{-(m_{1}-m_{2})x_{3}}  &  0\\
  0&  z-z_{2}&   \beta \,e^{-(m_{2}-m_{3})x_{3}}\\
  0&  0 &   z-z_{1}
\end{array}
\right)\,,
\label{eq:sepkinks}
\eeq
which corresponds to the rational map (\ref{eq:compmonopsep}), and gives the same result for the moduli space. 
\begin{eqnarray}
\mathcal M^{\rm mon}_{(1,1){\rm sep}}
&\equiv &\mathcal M^{\rm kink}_{(1,1){\rm sep}}
\end{eqnarray}
Let us consider now the case of coincident kinks. The most general configuration with at most two coincident vortices is described by the following moduli matrices. The first one is:
\beq
H^{u}_{(2,2){\rm align}}(z,x_{3})=
\left(
\begin{array}{ccc}
  1& 0  &  (a z + b)e^{-(m_{1}-m_{3})x_{3}} \\
  0&  1&   (c z+d)e^{-(m_{2}-m_{3})x_{3}} \\
  0&  0 &   (z-z_{0})^{2}
\end{array}
\right).
\eeq
Each vortex can support at most two kinks, thus the matrix above will generically describe a configuration with charges $(n_{1}^{\rm (k)},n_{2}^{\rm (k)})=(2,2)$. According to Eq.~(\ref{eq:kinknumber}), to have a configuration with charges $(1,1)$ we need some constraint on the moduli parameters such that, at positive infinity, the matrix above reduce to the following
\beq
H^{d}_{0}(z,x_{3})=
\left(
\begin{array}{ccc}
  z-z_{0} & 0  & 0  \\
 0 &  1 &   0 \\
  0 & 0  &   z-z_{0}
\end{array}
\right)\quad {\rm at} \quad x_{3}\rightarrow +\infty\,.
\eeq
This implies the existence of a neutral vortex. The constraint on the moduli parameters is thus the existence of a common factor $z-z_{0}$ on the same column. This enables us to reduce the degree of the polynomials by one, once we eliminate the neutral vortex:
\[
H^{u}_{(1,1){\rm align}}(z,x_{3})=\left(
\begin{array}{ccc}
 1 &  0 &  \gamma\, e^{-(m_{1}-m_{3})x_{3}} \\
  0 & 1  & \alpha\, e^{-(m_{2}-m_{3})x_{3}}  \\
  0 & 0  & z-z_{0}  
\end{array}
\right)\,.
\]
The same condition (\ref{eq:kinknumber}) implies $\gamma\neq0$. We may start from the following moduli matrix as well:
\beq
H^{u}_{(2,1){\rm align}}(z,x_{3})=
\left(
\begin{array}{ccc}
  1&   \alpha  \,e^{-(m_{1}-m_{2})x_{3}} &  \gamma \,e^{-(m_{1}-m_{3})x_{3}} \\
  0&  z-z_{0}&   \beta \,e^{-(m_{2}-m_{3})x_{3}} \\
  0&  0 &   z-z_{0}
\end{array}
\right).
\label{eq:kinkgen}
\eeq
It describes generic configurations of kinks with charges $(2,1)$. The condition to have a $(1,1)$ configuration is that the matrix above reduces to the following at positive infinity
\[
H^{d}_{0}(z,x_{3})=\left(
\begin{array}{ccc}
 z-z_{0} &  0 &  0\\
  0 & z-z_{0}  & 0 \\
  0 & 0  & 1 
\end{array}
\right), \quad {\rm at} \quad x_{3}\rightarrow \infty\,.
\]
Using  the $V$-equivalence explicitly, we can find the following conditions on the moduli parameters: $\alpha\,\beta=0$ and again $\gamma\neq0$. The case $\alpha=0$ reduces to the previous one (there is an additional neutral vortex in the second position). The case $\beta=0$ is new instead. The two cases differ by  the order of alignment of the kinks along the vortices\footnote{The order of kinks on a single vortex is fixed. To be able to invert this order, we need more coincident vortices to support the kinks.}. The case above correspond to the rational map  (\ref{eq:compmonop}). The observation above implies  the following:
\begin{eqnarray}
\mathcal M^{\rm mon}_{(1,1){\rm align}}
&\equiv &\mathcal M^{\rm kink}_{(1,1){\rm align}}
\end{eqnarray}

\subsubsection*{Minimal breaking: $U(3)\rightarrow U(2)\times U(1)$}
It is very simple to see what happens when we tune two masses to be equal $m_{1}=m_{2}$. Expression \ref{eq:sepkinks} reduce to 
\beq
H^{u}_{(1)}(z,x_{3})=
\left(
\begin{array}{ccc}
  1& \alpha  &  0\\
  0&  z-z_{2}&   \beta \,e^{-(m_{2}-m_{3})x_{3}}\\
  0&  0 &   z-z_{1}
\end{array}
\right)\,,
\eeq
which is nothing but a kink located at $z_{1}$ plus a neutral non-Abelian vortex located at $z_{2}$. Analogously to the monopole case, to obtain the most general kink we have to look at the matrix  (\ref{eq:kinkgen}). It the degenerate case, it corresponds to a neutral non-Abelian vortex plus a kink which can be described by the simplified matrix:
\beq
H^{u}_{(1)}(z,x_{3})=
\left(
\begin{array}{ccc}
  1&   0 &  \gamma \,e^{-(m_{1}-m_{3})x_{3}} \\
  0&  1&   \beta \,e^{-(m_{2}-m_{3})x_{3}} \\
  0&  0 &   z-z_{0}
\end{array}
\right).
\eeq
For the matrix above to correctly describe a kink, we are allowed to consider both the following asymptotic expression at positive infinity:
\[
H^{d}_{0}(z,x_{3})=\left(
\begin{array}{ccc}
 z-z_{0} &  0 &  0\\
  0 & 1  & 0 \\
  0 & 0  & 1 
\end{array}
\right), \quad \left(
\begin{array}{ccc}
 1 &  0 &  0\\
  0 & z-z_{0}  & 0 \\
  0 & 0  & 1 
\end{array}
\right), \quad {\rm at} \quad x_{3}\rightarrow \infty\,.
\]
This implies the condition $(\beta,\gamma)\neq(0,0)$, and a perfect matching with the rational map  (\ref{eq:compmonop2}) corresponding to a non-Abelian monopole:
\begin{eqnarray}
\mathcal M^{\rm mon}_{(1)}
&\equiv &\mathcal M^{\rm kink}_{(1)}
\end{eqnarray}

\section{Correspondence of Moduli Spaces} 
We have already discussed in detail the correspondence of fundamental and composite object in the $SU(3)$ case. In this section we use the rational map and the moduli matrix constructions for monopoles and kinks respectively, to show, in two additional examples, the isomorphism of the classical moduli spaces obtained by the two methods, when we correctly identify the monopole and kink numbers.


\subsection{Separated monopoles/kinks in $U(N)$ gauge theories}

\subsubsection*{Monopoles}

If we limit ourselves to study the moduli space of separate monopoles, we can describe it as a direct product of the moduli spaces of  single charge monopoles. This product should be then quotiented by permutations of identical monopoles\footnote{With identical monopoles we mean two objects with the same topological charge. The permutations exchange both spatial and internal (orientational) degree of freedom.}:
\beq
\mathcal M_{\rm sep}^{\rm mon}= \prod_{t=1}^{q-1}(\mathcal M^{\rm m}_{1_{t}}\times\dots\times \mathcal M^{\rm m}_{m_{t}})/\mathcal P_{m_{t}}\,,
\label{eq:permutmon}
\eeq
where $\mathcal M^{\rm m}_{i_{t}}$ is the moduli space of the $i$-th single monopole in the $t$-th charge sector.
The formula above can be considered as a consequence of the property of ``addition'' of rational maps with non-coincident poles \cite{BHMM}.
 
We can construct a fundamental monopole for a gauge theory with the generic symmetry breaking (\ref{eq:nonmaximal}) by embedding an Abelian monopole constructed from the breaking $SU(2)\rightarrow U(1)$ \cite{Ward:1982ft,Weinberg:1982ev,Auzzi:2004if,Bais:1997qy} into the larger gauge group. To construct a fundamental vortex, the broken $SU(2)$ must be chosen such that the $U(1)\subset SU(2)$ commutes with all the unbroken gauge symmetry, but not with two neighboring factors $S(U(n_{t})\times U(n_{t+1}))$. It is possible to show that this embedding results in a monopole configuration with a single non-vanishing topological charge $m_{t}=1$ related to the root $\vec\beta$ corresponding to the broken $SU(2)$. The holomorphic charges are also vanishing \cite{Bowman:1985kh}. The moduli space in this case is the product of the translational sector $\mathbb R^{3}$ times a  non-Abelian quotient which parametrize all the possible way to embed the Cartan subalgebra of the $SU(2)$ group into the product $S(U(n_{t})\times U(n_{t+1}))$ \cite{Bais:1997qy}:
\beq
\mathcal M_{(1),t}^{\rm mon}& =& \mathbb R^{3}\times S^{1}\ltimes \left(\frac{SU(n_{t})}{SU(n_{t}-1)\times U(1)}\times\frac{ SU(n_{t+1})}{SU(n_{t+1}-1)\times U(1)} \right) \nonumber \\
& =& \mathbb R^{3}\times S_{\rm el}^{1}\ltimes (\mathbb CP_{\rm mag}^{n_{t}-1}\times \mathbb CP_{\rm mag}^{n_{t+1}-1}).
\label{eq:singmonspace}
\eeq
The denominators in the quotients appear because they act trivially on the monopole configuration. Notice the existence of the non-trivial fibration of the magnetic orbit with the electric phase. 

\subsubsection*{Kinks}

Let us now switch to kinks. We show that the same expressions (\ref{eq:permutmon}) and (\ref{eq:singmonspace}) give  the moduli spaces of separated kinks. If we start from a configuration of well separated monopoles in the unbroken phase,  we can always choose the $x_{3}$-direction such that, once we enter the Higgs phase, there are no monopoles/kinks aligned on the same vortex. All the kinks will be separated in the transverse $z$ plane, and thus confined by non-coincident vortices. As was shown in Refs. \cite{Eto:2005yh,Eto:2006pg}, the moduli space of separated non-Abelian vortices splits into a symmetric product of single vortices. The same result trivially holds for kinks too:
 \beq
\mathcal M_{\rm sep}^{\rm kink}= \prod_{t=1}^{q-1}(\mathcal M^{k}_{1_{t}}\times\dots\times \mathcal M^{k}_{n_{t}^{\rm (k)}})/\mathcal P_{n_{t}^{\rm (k)}}\,,
\label{eq:permutkink}
\eeq
where $M^{k}_{i_{t}}$ is the moduli space of the $i$-th  non-Abelian kink in the $t$-th topological sector.  We now show that:
\beq
M^{k}_{i_{t}}\equiv M^{m}_{i_{t}}
\eeq
 
  To study a fundamental kink we just need a single vortex supporting it.  From (\ref{eq:kinknumber2}) we see that the presence of a single kink requires an orientational vector which has non-zero elements only in two neighboring  group of equal masses. This means that the fundamental kink can be constructed by embedding the kink present in the system\footnote{The embedding will not generate further moduli, because the kink will be invariant under the additional larger symmetries.}:
\beq
U(n_{t}+n_{t+1})\longrightarrow U(n_{t})\times U(n_{t+1}).
\eeq
A moduli matrix with the correct orientational vector is the following upper triangular sub-matrix:
\beq
H^{u}_{0}(z,x_{3})=\left(
\begin{array}{c|c|c}
 {\bf 1}_{n_{t}} &  0 & \vec b_{n_{t}} e^{-(m_{t+1}-m_{t})x_{3}}    \\
  \hline
  0  &    {\bf 1}_{n_{t+1}-1} & \vec c_{n_{t+1}-1} \\
  \hline
0   &      0   &  z-z_{0}  \\  
\end{array}
\right),
\eeq
where $\vec b_{n_{t}}$ and $\vec c_{n_{t+1}-1}$ are vectors of moduli of lengths  $n_{t}$ and $n_{t+1}-1$, respectively. 
At negative infinity we have
 \beq
H^{u}_{0}(z,x_{3})=\left(
\begin{array}{c|c|c}
 1_{n_{t}} & 0  & 0     \\
  \hline
 0 &    1_{n_{t+1}-1} & \vec c_{n_{t+1}-1} \\
  \hline
  0 &      0   &  z-z_{0}  \\  
\end{array}
\right), \quad x_{3}\rightarrow -\infty,
\eeq
which corresponds to a non-Abelian vortex with orientations $ \vec c_{n_{t+1}-1}$ for the gauge group factor $U(n_{t+1})$. The same matrix can be put into a lower triangular form using  the $V$-equivalence:
\beq
H^{d}_{0}(z,x_{3})=\left(
\begin{array}{c|c|c}
  z-z_{0}   &   0 & 0     \\
  \hline
  \vec  b'_{n_{t}-1}&  {\bf 1}_{n_{t}-1} & 0 \\
  \hline
  {\vec c\,'}_{n_{t+1}}e^{-(m_{t}-m_{t+1})x_{3}}& 0 & {\bf 1}_{n_{t+1}}  \\  
\end{array}
\right),
\eeq
where the new variables are related to the old one by:
\beq
b'_{n_{t}-1,t}=b_{n_{t},t+1}/b_{n_{t},1}, \quad c'_{n_{t}+1,t}=c_{n_{t},t}/b_{n_{t},1}, \quad c'_{n_{t}+1,n_{t}+1}=1/b_{n_{t},1}\,.
\eeq
The lower triangular matrix reduces at positive infinity to 
\beq
H^{d}_{0}(z,x_{3})=\left(
\begin{array}{c|c|c}
  z-z_{0}   &   0 &    0  \\
  \hline
  \vec  b'_{n_{t}-1}& {\bf 1}_{n_{t}-1} &  0\\
  \hline
  0&      0   & {\bf 1}_{n_{t+1}}  \\  
\end{array}
\right), \quad x_{3}\rightarrow +\infty, 
\eeq
which is a non-Abelian vortex constructed in the $U(n_{t})$ gauge group factor.
The moduli space of the non-Abelian kink is thus given by:
\begin{equation}
\mathcal M^{\rm kink}
=\mathbb {R}_{\rm vort}^{2}\times \mathbb {R}\times 
S_{\rm phase}^{1}\ltimes (\mathbb C P_{\rm vort}^{n_{t}-1}\times \mathbb C P_{\rm vort}^{n_{t+1}-1}).\label{eq:singkink}
\end{equation}
In the expression above, a factor of  $\mathbb {R}$ is given by the position $x_{3,0}$ of the kink along the $x_{3}$ line, which can be estimated by the following condition \cite{Eto:2006pg}:
\beq
|\vec b_{n_{t}}|^{2}e^{-2(m_{t+1}-m_{t})x_{3,0}}& \equiv & 1+|\vec c_{n_{t+1}-1}|^{2}, \quad {\rm or\,  equivalently } \nonumber \\
1+|\vec b'_{n_{t}-1}|^{2}  & \equiv &    |\vec c\,'_{n_{t+1}-1}|^{2}e^{-2(m_{t}-m_{t+1})x_{3,0}}.
\eeq
Notice that, as in the monopole case, the moduli space is completely given (excluding the translational moduli) by symmetries. {However there exists a difference:} while in the monopole case the moduli are generated by {\it gauge} symmetries,  in the kink case they are generated by the  color-flavor mixed {\it global} symmetries defined in Eq. (\ref{eq:colflav}) . 

{
The two $\mathbb CP$ factors in Eq.~(\ref{eq:singkink}) are not  part of the kink moduli space in the ordinary terminology but rather of the moduli spaces of the non-Abelian vortices attached to the monopole. However it is interesting to observe that they cannot be separated from a ``genuine" kink moduli ${\mathbb R} \times S^1_{\rm phase}$ because of the non-trivial fibering.
From the point of view of kinks, these $\mathbb CP$ factors are the boundary moduli which are non-normalizable. Therefore we simply call $\mathcal M^{\rm kink}$ in (\ref{eq:singkink}) as ``kink moduli" in this sense. The same structure was actually found for instantons trapped inside a non-Abelian vortex \cite{Eto:2004rz}.}\footnote{{The instantons can be regarded as sigma model lumps in the ${\mathbb C}P^{N-1}$ model on the vortex \cite{Eto:2004rz,Fujimori:2008ee}. The moduli space for a single instanton inside a single vortex in $U(2)$ gauge theory in the Higgs phase was found in \cite{Eto:2004rz} to be ${\mathcal M}^{\rm lump}=\mathbb C_{\rm vort} \times \mathbb C \times \mathbb C^* \ltimes {\mathbb C}P^1_{\rm vort}$. Here the ``genuine" lump moduli $\mathbb C \times \mathbb C^*$ is made of the position moduli $\mathbb C$ and the size and phase moduli $\mathbb C^*$ of the single lump. The latter part is fibered over the boundary moduli ${\mathbb C}P^1_{\rm vort}$ which is in fact the vortex orientational moduli.}}

{
Comparing Eqs.~(\ref{eq:singmonspace}) and (\ref{eq:singkink}), 
we notice a precise correspondence between the magnetic orbits of non-Abelian monopoles in unbroken phase and the orientational moduli spaces of non-Abelian vortices attached to the monopole from the both side of the $x_3$-direction 
in the Higgs phase.} 
That is, in the case of $SU(n_{1}+n_{2})\rightarrow SU(n_{1})\times SU(n_{2}) \times U(1)$ we have
\begin{eqnarray}
\mathcal M_{\rm mag}^{\rm mon} =  \mathcal M_{U(n_{1})\, {\rm vortex} }\times \mathcal M_{U(n_{2}) \, {\rm vortex}}.
 \end{eqnarray}
The relationship above extends the observation of  matching of non-Abelian fluxes made in \cite{Auzzi:2003fs,Auzzi:2003em,Eto:2006dx} to a one-to-one correspondence between non-Abelian monopole solutions and the non-Abelian  vortices that confine them.



\subsection{Minimal symmetry breaking}


\subsubsection*{Monopoles}
Let us analyze the  case of minimal symmetry breaking  $SU(N)\longrightarrow SU(N-1)\times U(1)$. The flag manifolds introduced in Eq. (\ref{eq:flagdef}) reduce to ordinary  projective spaces:
\beq
\mathbb F lag_{N-1,1}=\frac{SU(N)}{SU(N-1)\times U(1)}=\mathbb C P^{N-1}.
\eeq
There is only one topological magnetic charge $m$ characterizing the number of monopoles, and  $N-2$ holomorphic charges. A generic, based, rational map for the space $\mathbb C P^{N-1}$ can be constructed as a set of polynomials:
\beq
R_{m}& :& \mathbb C P^{1} \rightarrow \mathbb C P^{N-1}\,; \nonumber \\
R_{m}& = &  (P(z),Q_{1}(z),\dots,Q_{N-1}(z)), \quad \quad  {\rm deg }\, P(z)=m, \quad {\rm deg }\,Q_{i}(z)<m.
\eeq
The zeroes of the polynomial $P(z)$ can be regarded as the positions of the $m$ monopoles in the $z$ plane \cite{Manton:2004tk}, while the polynomials $Q_{i}(z)$, which cannot be all identically vanishing, include orientational degrees of freedom and  positions of the monopoles along the $x_{3}$ axis.

As already emphasized, a crucial point in the identification of monopoles with rational maps is the correspondence between a stratification of the monopole moduli space, in terms of  quantized holomorphic magnetic charges, and a stratification in the rational maps \cite{Murray:1989zk}. To give an explicit example, let us consider the case with two monopoles in $SU(3)$, with its corresponding stratification, following Ref. \cite{Bais:1997qy}. The value of the topological charge is $m=2$. In this case there are two possible values of the holomorphic charge $m_{\rm hol}=0,1$ \cite{Murray:1989zk}. The explicit rational map is:
\beq
R_{2}(z)=(z^{2}+\alpha z+\beta,a z + b, c z+d).
\eeq
The total moduli space has complex dimension 6.\footnote{The three polynomials do not have common factors, as a consequence at least one of the parameters $a,b,c,d$ must be non-zero.} A generic point in this moduli space belongs to the large 6-dimensional stratum $S_{m,m_{\rm hol}}=S_{2,1}$. Points belonging to the smaller stratum $S_{2,0}$ are defined by the following condition
\[
 {\rm Det} \left(
\begin{array}{cc}
 a & b   \\
 c & d 
\end{array}
\right)= {\rm Det} \,D \equiv 0.\label{eq:stratumcond}
\]
The small stratum has thus complex dimension 5. It corresponds to configurations which can be embedded into a smaller $\mathbb C P^{1}$ \cite{Bais:1997qy}. As a consequence, these configurations can be obtained as embedding of $SU(2)$ monopoles\footnote{The moduli space of two $SU(2)$ monopoles has complex dimension 4. The smaller stratum has an additional complex parameter which determines how we can embed $SU(2)$ into the full $SU(3)$.}.

\subsubsection*{Kinks}

Let us now determine the moduli space of kinks. A generic configuration containing $n^{\rm (k)}$ kinks confined by the same number of vortices in the minimally broken case is given by the following  $x_{3}$ dependent moduli matrix:
\beq
H^{u}_{0}(z,x_{3})=
\left(
\begin{array}{cccc}
  1& \cdots  &  Q_{1}(z)e^{-(m_{1}-m_{2})x_{3}} \\
   \vdots &  \ddots   &       \vdots         \\
  0&  \cdots&   Q_{N-1}e^{-(m_{1}-m_{2})x_{3}} \\
  0&  \cdots &   P(z)
\end{array}
\right) ,\quad {\rm deg }\, P(z)=n^{\rm (k)}, \quad {\rm deg }\,Q_{i}(z)<n^{\rm (k)}.
\label{eq:modmatminimal}
\eeq
It is simple to check that, as soon as the $Q_{i}(z)$'s are not all vanishing, it is possible to put the matrix above in a lower triangular form:
\beq
H^{d}_{0}(z,x_{3})=
\left(
\begin{array}{ccccc}
  1& \cdots  &  0 & \dots & 0\\
   \vdots &  \ddots   &       \vdots        & \cdots  & \vdots\\
  0&  0 &   P(z) &  0  & \vdots \\
   \vdots &  \cdots   &       \vdots        & \ddots  & \vdots\\
  0&  0&   Q'_{N-1}e^{-(m_{2}-m_{1})x_{3}} & \dots & 1
\end{array}
\right),
\eeq
where the polynomials can lie in one of the first $N-1$ columns. In terms of the analysis of the previous section, this ensures the presence of $n^{\rm (k)}$ kinks. The matrix (\ref{eq:modmatminimal}) exactly corresponds to the rational map describing non-Abelian monopoles in the minimal breaking case. 

Let us work out explicitly the $SU(3)$ case with two kinks. The moduli matrix is
\beq
H_{0}(z,x_{3})=
\left(
\begin{array}{ccc}
  1& 0  &  (a z + b)e^{-(m_{1}-m_{2})x_{3}} \\
  0&  1&   (c z+d)e^{-(m_{1}-m_{2})x_{3}} \\
  0&  0 &   z^{2}+\alpha z+\beta
\end{array}
\right).
\label{eq:uppersu3}
\eeq
It correctly represents a double kink, confined by two vortices centered along the zeros of the polynomial $z^{2}+\alpha z+\beta$. It is also important to identify a concept of stratification for the moduli space of kinks, as defined by the moduli matrix. In the case at hand, the larger stratum is defined by the set of points in the moduli space for which the matrix (\ref{eq:uppersu3}) can be put into the following lower triangular form at positive infinity:
\beq
H_{0}(z,+\infty)=
\left(
\begin{array}{ccc}
  z-\phi & \eta  & 0 \\
 \tilde \eta &  z-\tilde\phi  &   0 \\
  0 & 0  &   1
\end{array}
\right).
\label{eq:largekink}
\eeq
Kinks corresponding to the case above correspond to monopoles that have a non vanishing holomorphic charge $n_{\rm hol}^{\rm (k)}=1$\footnote{One can add a small mass difference for the first two flavors to transform the holomorphic charge into a topological one. If we calculate the value of this second topological  charge in the case  (\ref{eq:largekink}) using the formula given in the previous section we find $n_{2}^{\rm (k)}=1$.}. The condition on the moduli parameters which gives the smaller stratum is then the same as  equation (\ref{eq:stratumcond}). In fact, as was shown in Ref. \cite{Eto:2006cx}, the condition ${\rm Det}\, D=0$ on the moduli matrix parameters implies parallel vortices which can be embedded, together with the kinks that they support, into an $SU(2)$ subgroup.

We notice here that the stratification of the composite $SU(3)$ monopole considered in this section corresponds to the stratification of the moduli space of two composite $U(2)$ non-Abelian vortices\footnote{These vortices are exactly the ones required to confine the composite monopole in the Higgs phase.} originally considered in Ref. \cite{Eto:2006cx,Eto:2006dx,Eto:2006db}. In those works, the moduli space of non-Abelian vortices was decomposed into submanifold and it was proposed to associate each stratum to different representations of the color-flavor symmetry group. In the case at hand, for example, two composite vortices associated with a fundamental representation of the $SU(2)_{C+F}$ group form a composite state with a moduli space which has two strata, associated respectively with the singlet and the triplet representations. This issue was clarified in the general case in Ref. \cite{Eto:2010aj}, where the moduli space of $k$ composite $U(N)$ non-Abelian vortices  is decomposed into strata associated to all the $k$-tensor representations of $SU(N)$. The K\"ahler potential in each of this stratum was also computed, and it turns out to be proportional to integer quantities. The size of these strata can be compared with a similar property of the monopole strata \cite{Murray:1989zk}.


%
%
%
%


\section{Equivalence of the Moduli Matrix and the Rational Map Construction} 
\label{sec:equiv}
In the previous sections we have shown, in some special cases, that the moduli spaces of monopoles and kinks are isomorphic. It is plausible that these results hold in the most general case, with arbitrary number of monopoles and generic symmetry breaking pattern. The physical reason behind this expectation is that, as already emphasized, monopoles and kinks are essentially the same objects in two different phases of the theory. Supported by the non-trivial checks we considered in this paper, we are led to state the following:
\emph{
\subsubsection*{There is a one-to-one correspondence between:}
\begin{enumerate}
\item The moduli space of non-Abelian monopoles, or equivalently,\\
 The moduli space of based rational maps into flag manifolds;
\item The moduli space of non-Abelian kinks, or equivalently,\\ 
       The set of moduli matrices modulo $V$-equivalence;
\end{enumerate}
\vspace{.1cm}
provided that we eliminate, from the kink-vortex system, neutral vortices.}
\vspace{.3cm}

\subsection{Hitchin and Bogomol'nyi equations}
In this section we give a sketch of a proof for the correspondence. The technical link between the moduli matrix for kinks and the rational  map for monopoles  is  the Hitchin operator (\ref{eq:hitchin}). It appears naturally in the second of the Bogomol'nyi equations (\ref{eq:kinkbps}). In fact, after the field redefinition (\ref{eq:fieldred}), the following equation
\begin{equation}
\nabla_{3}Q +  \Phi Q+  Q  M =0 
\end{equation}
reduces exactly to the Hitchin equation for the field $\Phi$:
\begin{equation}
(\nabla_{3} +  \Phi) \Psi =0, \quad Q=\Psi P\,,
\end{equation}
which must be solved with the following boundary condition:
\beq
|Q|=\sqrt\xi \quad \Rightarrow \quad |\Psi| = \sqrt \xi \,P^{-1}.
\label{eq:psibound}
\eeq
If we recall that we constructed the rational maps for monopoles by studying the scattering coefficient of the auxiliary Hitchin equation as an operator on  an auxiliary field $\psi$ (Eq. (\ref{eq:hitchin})), we see that the construction of kinks reduces ``almost'' exactly to the same problem. Moreover, putting monopoles in the Higgs phase gives a simple physical interpretation for the rational map construction:
\begin{itemize}
\item The Hitchin equation naturally arises  as part of the Bogomol'nyi equations for kinks.
\item  The auxiliary field $\psi$ is provided by the matter fields  $\Psi$ introduced in order to enter the Higgs phase. 
\item The arbitrary direction $x_{3}$ is  the direction of formation of vortices.
\end{itemize}

There is, however, a difference when we analyze the Hitchin equation in the unbroken phase and in the Higgs phase. In the unbroken phase the Higgs field $\Phi$ has a polynomial tail which is proportional to the magnetic charges (see equation (\ref{eq:higgsboundary})). In the Higgs phase it has an exponential decay, determined by the Fayet-Iliopoulos term.
The information about the topological charges of kinks must be determined, in this case, by comparing the vortex numbers at both infinities. 

Notice, however,  that the construction for kinks holds for arbitrary small values of the Fayet-Iliopolous term $\sqrt\xi$. In the regime $\sqrt\xi\ll|m_{i}-m_{j}|$, monopoles are weakly confined (the width of vortices is much larger than the size of monopoles), and it looks like that they are still in the unbroken phase. For very small $\xi$, indeed, the values of the matter fields are very small, and the Bogomol'nyi equations for kinks reduce, at first order, to those for  free monopoles. In this regime we can then ignore the backreaction of the matter fields on the monopole, and   directly apply the rational map construction for monopoles. The BPS equation for the matter fields, then, can be considered as the Hitchin equation on a fixed background, exactly as needed to interpret $\Psi$ as the scattering fields of the rational map construction. We can then rely on the fact that the parameter $\xi$ cannot change the dimension of the moduli space to claim the validity of the analysis at large $\xi$.

\section{Discussion}

We have studied the precise correspondence between 
the moduli spaces of monopoles in the unbroken phase and 
in the Higgs phase, including non-Abelian monopoles.
The former described in the rational map construction 
has been found to coincide with the latter
described by the moduli matrix formalism. 
Nontrivial fiber structure of an electric orbit 
over magnetic orbits in the moduli space of 
non-Abelian monopoles in the unbroken phase becomes 
the kink moduli fibered over the moduli of a non-Abelian vortex.
We thus have found that the moduli space of monopoles coincides with 
the moduli space of kinks if we include 
the boundary moduli which are in fact the moduli of vortices 
attached to kinks.

\changed{
In this paper we have studied monopoles in $SU(N)$ or $U(N)$ gauge theories.
Changing gauge groups is one interesting extension. 
Especially the $SO$ and $USp$ cases have been studied for monopoles and 
also recently for vortices  in 
\cite{Ferretti:2007rp,Eto:2008qw,Eto:2009bg,Gudnason:2010jq,Gudnason:2010rm},
and there will be a similar relation between monopoles in 
unbroken and Higgs phases for $SO$ and $USp$ gauge theories.
The case of arbitrary gauge groups \cite{Eto:2008yi} will be possible 
in principle.
}

The correspondence of monopoles and kinks studied in this paper 
can be extended to the one of Yang-Mills instantons and lumps.
This is because in the Higgs phase instantons can stably exist inside a non-Abelian vortex \cite{Eto:2004rz,Fujimori:2008ee}. 
Such trapped instantons can be regarded as lumps in the ${\mathbb C}P^{N-1}$ model. 
For the case of a single vortex, a similar correspondence can be 
understood from the work of Atiyah \cite{Atiyah:1984tk}; 
The moduli space of $SO(2)$ invariant $SU(2)$ Yang-Mills instantons, 
placed on a plane, is isomorphic to a space of a rational map into ${\mathbb C}P^{1}$. Extension to the case of instantons is surely interesting for instance in application to the instanton counting \cite{Nekrasov:2002qd}.
It is well known that 
monopoles and instantons are related by the Nahm transformation \cite{Nahm}
or T-duality in the corresponding D-brane configuration. The same relation should hold when we enter the Higgs phase. In fact the T-duality between 
domain walls and vortices on a cylinder \cite{Eto:2006mz}  
and on a torus \cite{Eto:2007aw} was found already, 
and such a relation was found to hold for kinks and lumps inside 
a non-Abelian vortex \cite{Eto:2004rz}.

\section*{Acknowledgments}

W.V. would like to thank M. Shifman and Kenichi Konishi for their valuable comments on the preliminary version of the paper. 
The work of M.N. is partially supported by a Grant-in-Aid for Scientific Research No. 20740141 from 
the Ministry of Education, Culture, Sports, Science and Technology, Japan.
The work of W.V. is supported by the DOE grant DE-FG02-94ER40823.
\appendix

\section{Scattering data}
The analysis of  section \ref{sec:equiv} suggests, indeed, that the moduli matrix can be considered as an explicit realization of the scattering data of the Hitchin equation in the background of non-Abelian monopoles. 

To reconstruct the moduli matrix,  we  simply patch together $N$ independent  solutions of the Hitchin operator,
\begin{equation}
\Psi_{\rm scatt}=(\psi_{1},\dots,\psi_{i},\dots,\psi_{N})\,,
\end{equation}
which converge at negative infinity. We can chose, for example, the following set:
\begin{equation}
\psi_{i}\sim  \left(
\begin{array}{c}
  0 \\
   \vdots  \\
   1 \\
   0  \\
   \vdots
\end{array}
\right)e^{m_{i} x_{3}}, \quad x_{3}\rightarrow  - \infty.
\label{eq:reconstruct}
\end{equation}
The single non-zero field is in the $i$-th position. As already explained, we can track these solutions toward large positive values of $x_{3}$, and rewrite them in term of scattering data. The condition of holomorphicity  of these data holds here too at large $|x_{3}|$ values.  At positive infinity, the $\Psi$ matrix reads:
\beq
\Psi_{\rm scatt}
\sim e^{M  x_{3}}
\left(
\begin{array}{ccccc}
 1_{1} & \dots  &  R_{1,i}(z) &  \dots   &  R_{1,N}(z) \\
  \vdots & \ddots  &   &  & \vdots \\
 R_{i,1}(z)   &    &  1_{i} &  & R_{i,N}(z) \\
    \vdots & & & \ddots & \vdots \\
  R_{N,1}(z)   & \dots &  R_{N,i}(z)   & \dots &  1_{N}
  \end{array}
\right), \quad x_{3}\rightarrow  +\infty,
\eeq
where the $R_{i,j}(z)$ are rational holomorphic functions which tend to zero at large $z$. Furthermore, we may assume that all scattering coefficients $R_{i,j}$ of the same solution $\psi_{j}$ are continuous up to a finite number $n_{i}$ of points, the zeroes of the polynomial $P_{j}(z)$:
\beq
R_{i,j}\equiv Q_{i,j}(z)/P_{j}(z).
\eeq
If we now  fix the normalization of $\Psi_{\rm scatt}$ using the boundary equation (\ref{eq:psibound}), we can directly identify $\Psi_{\rm scatt}$ with $\Psi$:
\beq
\Psi=\Psi_{\rm scatt}\cdot {\rm diag} (P_{1}(z),\dots,P_{N}(z))
\eeq
Given the relation $|\Psi|=|S^{-1}||H_{0}|$ we obtain $H_{0}$ and $S$ from $\Psi_{\rm scatt}$. 
\begin{eqnarray}
H_{0}\equiv
\left(
\begin{array}{ccccc}
 P_{1} & \dots  &  Q_{1,i}(z) &  \dots   &  Q_{1,N}(z) \\
  \vdots & \ddots  &   &  & \vdots \\
 Q_{i,1}(z)   &    &  P_{i} &  & Q_{i,N}(z) \\
    \vdots & & & \ddots & \vdots \\
  Q_{N,1}(z)   & \dots &  Q_{N,i}(z)   & \dots &  P_{N}
  \end{array}
\right),\quad S\rightarrow e^{-Mx_{3}}, \quad |x_{3}| \rightarrow\infty,
\end{eqnarray}
The moduli matrix above is already in a form where the $V$-equivalence is completely fixed, and all the coefficients are true moduli of the configuration. If one wishes, one can put the matrices in an upper triangular form, to make full contact with the discussions in the bulk of the paper.


%
%
%
%
%
%
%
%
%
%
%
%
%
%
%
%
%
%
%
%
%


\bibliography{Bibliographysmart}

\begin{thebibliography}{100}
\ifx\href\asklfhas\newcommand{\href}[2]{#2}\fi
\ifx\arxivref\asklfhas\newcommand{\arxivref}[1]{\href{http://arxiv.org/abs/#1}%
{#1}}\fi
\ifx\doiref\asklfhas\newcommand{\doiref}[2]{\href{http://dx.doi.org/#1}{#2}}\fi
\raggedright
\small
\parskip 0pt

\bibitem{Dirac:1931kp}
P.~A.~M.~Dirac,
\textit{``{Quantised singularities in the electromagnetic field}''},
\textsf{Proc.~Roy.~Soc.~Lond.~A133,~60~(1931)}.
%
\bibitem{'tHooft:1974qc}
G.~'t~Hooft,
\textit{``{Magnetic Monopoles in Unified Gauge Theories}''},
\textsf{\doiref{10.1016/0550-3213(74)90486-6}{Nucl.~Phys.~B79,~276~(1974)}}.
%
\bibitem{Polyakov:1974ek}
A.~M.~Polyakov,
\textit{``{Particle spectrum in quantum field theory}''},
\textsf{JETP~Lett.~20,~194~(1974)}.
%
\bibitem{Preskill:1979zi}
J.~Preskill,
\textit{``{Cosmological Production of Superheavy Magnetic Monopoles}''},
\textsf{\doiref{10.1103/PhysRevLett.43.1365}{Phys.~Rev.~Lett.~43,~1365~(1979)}%
}.
%
\bibitem{Zeldovich:1978wj}
Y.~B.~Zeldovich and M.~Y.~Khlopov,
\textit{``{On the Concentration of Relic Magnetic Monopoles in the
  Universe}''},
\textsf{\doiref{10.1016/0370-2693(78)90232-0}{Phys.~Lett.~B79,~239~(1978)}}.
%
\bibitem{Sato:1980yn}
K.~Sato,
\textit{``{First Order Phase Transition of a Vacuum and Expansion of the
  Universe}''},
\textsf{Mon.~Not.~Roy.~Astron.~Soc.~195,~467~(1981)}.
%
\bibitem{Guth:1980zm}
A.~H.~Guth,
\textit{``{The Inflationary Universe: A Possible Solution to the Horizon and
  Flatness Problems}''},
\textsf{\doiref{10.1103/PhysRevD.23.347}{Phys.~Rev.~D23,~347~(1981)}}.
%
\bibitem{'tHooft:1981ht}
G.~'t~Hooft,
\textit{``{Topology of the Gauge Condition and New Confinement Phases in
  Nonabelian Gauge Theories}''},
\textsf{\doiref{10.1016/0550-3213(81)90442-9}{Nucl.~Phys.~B190,~455~(1981)}}.
%
\bibitem{Mandelstam:1974pi}
S.~Mandelstam,
\textit{``{Vortices and Quark Confinement in Nonabelian Gauge Theories}''},
\textsf{\doiref{10.1016/0370-1573(76)90043-0}{Phys.~Rept.~23,~245~(1976)}}.
%
\bibitem{Bogomolny:1975de}
E.~B.~Bogomolny,
\textit{``{Stability of Classical Solutions}''},
\textsf{Sov.~J.~Nucl.~Phys.~24,~449~(1976)}.
%
\bibitem{Prasad:1975kr}
M.~Prasad and C.~M.~Sommerfield,
\textit{``{An Exact Classical Solution for the 't Hooft Monopole and the
  Julia-Zee Dyon}''},
\textsf{\doiref{10.1103/PhysRevLett.35.760}{Phys.Rev.Lett.~35,~760~(1975)}}.
%
\bibitem{Seiberg:1994rs}
N.~Seiberg and E.~Witten,
\textit{``{Monopole Condensation, And Confinement In N=2 Supersymmetric
  Yang-Mills Theory}''},
\textsf{\doiref{10.1016/0550-3213(94)90124-4}{Nucl.~Phys.~B426,~19~(1994)}},
\texttt{\arxivref{hep-th/9407087}}.
%
\bibitem{Seiberg:1994aj}
N.~Seiberg and E.~Witten,
\textit{``{Monopoles, duality and chiral symmetry breaking in N=2
  supersymmetric QCD}''},
\textsf{\doiref{10.1016/0550-3213(94)90214-3}{Nucl.~Phys.~B431,~484~(1994)}},
\texttt{\arxivref{hep-th/9408099}}.
%
\bibitem{Goddard:1976qe}
P.~Goddard, J.~Nuyts and D.~I.~Olive,
\textit{``{Gauge Theories and Magnetic Charge}''},
\textsf{\doiref{10.1016/0550-3213(77)90221-8}{Nucl.~Phys.~B125,~1~(1977)}}.
%
\bibitem{Montonen:1977sn}
C.~Montonen and D.~I.~Olive,
\textit{``{Magnetic Monopoles as Gauge Particles?}''},
\textsf{\doiref{10.1016/0370-2693(77)90076-4}{Phys.~Lett.~B72,~117~(1977)}}.
%
\bibitem{Seiberg:1994pq}
N.~Seiberg,
\textit{``{Electric - magnetic duality in supersymmetric nonAbelian gauge
  theories}''},
\textsf{\doiref{10.1016/0550-3213(94)00023-8}{Nucl.~Phys.~B435,~129~(1995)}},
\texttt{\arxivref{hep-th/9411149}}.
%
\bibitem{Sen:1994yi}
A.~Sen,
\textit{``{Dyon - monopole bound states, selfdual harmonic forms on the multi -
  monopole moduli space, and SL(2,Z) invariance in string theory}''},
\textsf{\doiref{10.1016/0370-2693(94)90763-3}{Phys.~Lett.~B329,~217~(1994)}},
\texttt{\arxivref{hep-th/9402032}}.
%
\bibitem{Carlino:2000ff}
G.~Carlino, K.~Konishi and H.~Murayama,
\textit{``{Dynamics of supersymmetric SU(n(c)) and USp(2n(c)) gauge
  theories}''},
\textsf{JHEP~0002,~004~(2000)},
\texttt{\arxivref{hep-th/0001036}}.
%
\bibitem{Carlino:2000uk}
G.~Carlino, K.~Konishi and H.~Murayama,
\textit{``{Dynamical symmetry breaking in supersymmetric SU(n(c)) and
  USp(2n(c)) gauge theories}''},
\textsf{\doiref{10.1016/S0550-3213(00)00482-X}{Nucl.~Phys.~B590,~37~(2000)}},
\texttt{\arxivref{hep-th/0005076}}.
%
\bibitem{Ward:1982ft}
R.~Ward,
\textit{``{Deformations of the Embedding of the SU(2) Monopole Solution in
  SU(3)}''},
\textsf{\doiref{10.1007/BF01212178}{Commun.Math.Phys.~86,~437~(1982)}}.
%
\bibitem{Weinberg:1982ev}
E.~J.~Weinberg,
\textit{``{Fundamental Monopoles in Theories With Arbitrary Symmetry
  Breaking}''},
\textsf{\doiref{10.1016/0550-3213(82)90324-8}{Nucl.~Phys.~B203,~445~(1982)}}.
%
\bibitem{Auzzi:2004if}
R.~Auzzi, S.~Bolognesi, J.~Evslin, K.~Konishi and H.~Murayama,
\textit{``{NonAbelian monopoles}''},
\textsf{\doiref{10.1016/j.nuclphysb.2004.08.041}{Nucl.~Phys.~B701,~207~(2004)}%
},
\texttt{\arxivref{hep-th/0405070}}.
%
\bibitem{Abouelsaood:1982dz}
A.~Abouelsaood,
\textit{``{Are There Chromodyons?}''},
\textsf{\doiref{10.1016/0550-3213(83)90195-5}{Nucl.~Phys.~B226,~309~(1983)}}.
%
\bibitem{Abouelsaood:1983gw}
A.~Abouelsaood,
\textit{``{Chromodyons and equivariant gauge transformations}''},
\textsf{\doiref{10.1016/0370-2693(83)91327-8}{Phys.~Lett.~B125,~467~(1983)}}.
%
\bibitem{Nelson:1983bu}
P.~C.~Nelson and A.~Manohar,
\textit{``{Global Color Is Not Always Defined}''},
\textsf{\doiref{10.1103/PhysRevLett.50.943}{Phys.~Rev.~Lett.~50,~943~(1983)}}.
%
\bibitem{Balachandran:1982gt}
A.~P.~Balachandran et~al.,
\textit{``{Monopole Topology and the Problem of Color}''},
\textsf{\doiref{10.1103/PhysRevLett.50.1553}{Phys.~Rev.~Lett.~50,~1553~(1983)}%
}.
%
\bibitem{Horvathy:1984yg}
P.~A.~Horvathy and J.~H.~Rawnsley,
\textit{``{Internal Symmetries Of Nonabelian Gauge Field Configurations}''},
\textsf{\doiref{10.1103/PhysRevD.32.968}{Phys.~Rev.~D32,~968~(1985)}}.
%
\bibitem{Horvathy:1985bp}
P.~A.~Horvathy and J.~H.~Rawnsley,
\textit{``{The Problem Of 'Global Color' In Gauge Theories}''},
\textsf{\doiref{10.1063/1.527119}{J.~Math.~Phys.~27,~982~(1986)}}.
%
\bibitem{Horvathy:1988ge}
P.~A.~Horvathy, L.~O'Raifeartaigh and J.~H.~Rawnsley,
\textit{``{Monopole charge instability}''},
\textsf{\doiref{10.1142/S0217751X88000291}{Int.~J.~Mod.~Phys.~A3,~665~(1988)}},
\texttt{\arxivref{0909.2523}}.
%
\bibitem{Nelson:1983fn}
P.~C.~Nelson and S.~R.~Coleman,
\textit{``{What Becomes of Global Color}''},
\textsf{\doiref{10.1016/0550-3213(84)90013-0}{Nucl.~Phys.~B237,~1~(1984)}}.
%
\bibitem{Dorey:1995me}
N.~Dorey, C.~Fraser, T.~J.~Hollowood and M.~A.~C.~Kneipp,
\textit{``{Non-abelian duality in N=4 supersymmetric gauge theories}''},
\texttt{\arxivref{hep-th/9512116}}.
%
\bibitem{Dorey:1996jh}
N.~Dorey, C.~Fraser, T.~J.~Hollowood and M.~A.~C.~Kneipp,
\textit{``{S-duality in N=4 supersymmetric gauge theories}''},
\textsf{\doiref{10.1016/0370-2693(96)00773-3}{Phys.~Lett.~B383,~422~(1996)}},
\texttt{\arxivref{hep-th/9605069}}.
%
\bibitem{Bais:1997qy}
F.~A.~Bais and B.~J.~Schroers,
\textit{``{Quantisation of monopoles with non-abelian magnetic charge}''},
\textsf{\doiref{10.1016/S0550-3213(97)00778-5}{Nucl.~Phys.~B512,~250~(1998)}},
\texttt{\arxivref{hep-th/9708004}}.
%
\bibitem{Nahm}
W.~Nahm,
\textit{``{The Construction of All Self-dual Multi-Monopoles by The ADHM
  Method}, Monopoles in Quantum Fiels Theories, Singapore, World Scientific,
  1982''}.
%
\bibitem{Atiyah:1978ri}
M.~F.~Atiyah, N.~J.~Hitchin, V.~G.~Drinfeld and Y.~I.~Manin,
\textit{``{Construction of instantons}''},
\textsf{\doiref{10.1016/0375-9601(78)90141-X}{Phys.~Lett.~A65,~185~(1978)}}.
%
\bibitem{Hitchin:1982gh}
N.~J.~Hitchin,
\textit{``{Monopoles and Geodesics}''},
\textsf{\doiref{10.1007/BF01208717}{Commun.~Math.~Phys.~83,~579~(1982)}}.
%
\bibitem{Ward:1977ta}
R.~S.~Ward,
\textit{``{On Selfdual gauge fields}''},
\textsf{\doiref{10.1016/0375-9601(77)90842-8}{Phys.~Lett.~A61,~81~(1977)}}.
%
\bibitem{Forgacs:1980ym}
P.~Forgacs, Z.~Horvath and L.~Palla,
\textit{``{Exact Multi - Monopole Solutions in the Bogomolny-Prasad-
  Sommerfield Limit}''},
\textsf{\doiref{10.1016/0370-2693(81)91115-1}{Phys.~Lett.~B99,~232~(1981)}}.
%
\bibitem{Donaldson:1985id}
S.~K.~Donaldson,
\textit{``{Nahm's Equations And The Classification Of Monopoles}''},
\textsf{\doiref{10.1007/BF01214583}{Commun.~Math.~Phys.~96,~387~(1984)}}.
%
\bibitem{Jarvis}
S.~Jarvis,
\textit{``{Euclidean monopoles and rational maps}''},
\textsf{Proc.~London~Math.~Soc.~77,~170~(1998)}.
%
\bibitem{Jarvis2}
S.~Jarvis,
\textit{``{Construction Of Euclidean Monopoles}''},
\textsf{Proc.~London~Math.~Soc.~77,~193~(1998)}.
%
\bibitem{Dancer}
A.~Dancer,
\textit{``{A family of Hyperkahler Manifolds}''},
\textsf{Quart.~Jour.~Math.~45,~463~(1994)}.
%
\bibitem{Murray:1989zk}
M.~Murray,
\textit{``{Stratifying Monopoles and Rational Maps}''},
\textsf{\doiref{10.1007/BF01228347}{Commun.~Math.~Phys.~125,~661~(1989)}}.
%
\bibitem{Tong:2003pz}
D.~Tong,
\textit{``{Monopoles in the Higgs phase}''},
\textsf{\doiref{10.1103/PhysRevD.69.065003}{Phys.~Rev.~D69,~065003~(2004)}},
\texttt{\arxivref{hep-th/0307302}}.
%
\bibitem{Auzzi:2003fs}
R.~Auzzi, S.~Bolognesi, J.~Evslin, K.~Konishi and A.~Yung,
\textit{``{Nonabelian superconductors: Vortices and confinement in N = 2
  SQCD}''},
\textsf{\doiref{10.1016/j.nuclphysb.2003.09.029}{Nucl.~Phys.~B673,~187~(2003)}%
},
\texttt{\arxivref{hep-th/0307287}}.
%
\bibitem{Hindmarsh:1985xc}
M.~Hindmarsh and T.~W.~B.~Kibble,
\textit{``{Beads on Strings}''},
\textsf{\doiref{10.1103/PhysRevLett.55.2398}{Phys.~Rev.~Lett.~55,~2398~(1985)}%
}.
%
\bibitem{Hanany:2004ea}
A.~Hanany and D.~Tong,
\textit{``{Vortex strings and four-dimensional gauge dynamics}''},
\textsf{JHEP~0404,~066~(2004)},
\texttt{\arxivref{hep-th/0403158}}.
%
\bibitem{Shifman:2004dr}
M.~Shifman and A.~Yung,
\textit{``{Non-Abelian string junctions as confined monopoles}''},
\textsf{\doiref{10.1103/PhysRevD.70.045004}{Phys.~Rev.~D70,~045004~(2004)}},
\texttt{\arxivref{hep-th/0403149}}.
%
\bibitem{Dorey:1998yh}
N.~Dorey,
\textit{``{The BPS spectra of two-dimensional supersymmetric gauge theories
  with twisted mass terms}''},
\textsf{JHEP~9811,~005~(1998)},
\texttt{\arxivref{hep-th/9806056}}.
%
\bibitem{Dorey:1999zk}
N.~Dorey, T.~J.~Hollowood and D.~Tong,
\textit{``{The BPS spectra of gauge theories in two and four dimensions}''},
\textsf{JHEP~9905,~006~(1999)},
\texttt{\arxivref{hep-th/9902134}}.
%
\bibitem{Abraham:1992vb}
E.~R.~C.~Abraham and P.~K.~Townsend,
\textit{``{Q kinks}''},
\textsf{\doiref{10.1016/0370-2693(92)90122-K}{Phys.~Lett.~B291,~85~(1992)}}.
%
\bibitem{Abraham:1992qv}
E.~R.~C.~Abraham and P.~K.~Townsend,
\textit{``{More on Q kinks: A (1+1)-dimensional analog of dyons}''},
\textsf{\doiref{10.1016/0370-2693(92)91558-Q}{Phys.~Lett.~B295,~225~(1992)}}.
%
\bibitem{Hanany:2005bq}
A.~Hanany and D.~Tong,
\textit{``{On monopoles and domain walls}''},
\textsf{\doiref{10.1007/s00220-006-0056-7}{Commun.~Math.~Phys.~266,~647~(2006)%
}},
\texttt{\arxivref{hep-th/0507140}}.
%
\bibitem{Eto:2008dm}
M.~Eto, T.~Fujimori, M.~Nitta, K.~Ohashi and N.~Sakai,
\textit{``{Domain Walls with Non-Abelian Clouds}''},
\textsf{\doiref{10.1103/PhysRevD.77.125008}{Phys.~Rev.~D77,~125008~(2008)}},
\texttt{\arxivref{0802.3135}}.
%
\bibitem{Eto:2006pg}
M.~Eto, Y.~Isozumi, M.~Nitta, K.~Ohashi and N.~Sakai,
\textit{``{Solitons in the Higgs phase: The moduli matrix approach}''},
\textsf{J.~Phys.~A39,~R315~(2006)},
\texttt{\arxivref{hep-th/0602170}}.
%
\bibitem{Bais:1978yh}
F.~A.~Bais and H.~A.~Weldon,
\textit{``{Exact Monopole Solutions In SU(N) Gauge Theory}''},
\textsf{\doiref{10.1103/PhysRevLett.41.601}{Phys.~Rev.~Lett.~41,~601~(1978)}}.
%
\bibitem{Weinberg:1982jh}
E.~J.~Weinberg,
\textit{``{A Continuous Family Of Magnetic Monopole Solutions}''},
\textsf{\doiref{10.1016/0370-2693(82)90265-9}{Phys.Lett.~B119,~151~(1982)}}.
%
\bibitem{Ward:1981qq}
R.~S.~Ward,
\textit{``{Magnetic Monopoles with Gauge Group SU(3) Broken to U(2)}''},
\textsf{\doiref{10.1016/0370-2693(81)90831-5}{Phys.~Lett.~B107,~281~(1981)}}.
%
\bibitem{Weinberg:1979zt}
E.~J.~Weinberg,
\textit{``{Fundamental Monopoles and Multi-Monopole Solutions for Arbitrary
  Simple Gauge Groups}''},
\textsf{\doiref{10.1016/0550-3213(80)90245-X}{Nucl.~Phys.~B167,~500~(1980)}}.
%
\bibitem{Englert:1976ng}
F.~Englert and P.~Windey,
\textit{``{Quantization Condition for 't Hooft Monopoles in Compact Simple Lie
  Groups}''},
\textsf{\doiref{10.1103/PhysRevD.14.2728}{Phys.~Rev.~D14,~2728~(1976)}}.
%
\bibitem{Julia:1975ff}
B.~Julia and A.~Zee,
\textit{``{Poles with Both Magnetic and Electric Charges in Nonabelian Gauge
  Theory}''},
\textsf{\doiref{10.1103/PhysRevD.11.2227}{Phys.~Rev.~D11,~2227~(1975)}}.
%
\bibitem{BHMM}
C.~Boyer, B.~Mann, J.~Hurtubise and R.~Milgram,
\textit{``{The topology of the space of rational maps into generalized flag
  manifolds}''},
\textsf{Acta~Mathematica~173,~61~(1994)}.
%
\bibitem{Manton:2004tk}
N.~S.~Manton and P.~Sutcliffe,
\textit{``{Topological solitons}''},
Cambridge, UK: Univ. Pr. (2004) 493 p.
%
\bibitem{Hurtubise:1985vq}
J.~Hurtubise,
\textit{``{Monopoles And Rational Maps: A Note On A Theorem Of Donaldson}''},
\textsf{\doiref{10.1007/BF01212447}{Commun.~Math.~Phys.~100,~191~(1985)}}.
%
\bibitem{Kampmeijer:2008wz}
L.~Kampmeijer, J.~K.~Slingerland, B.~J.~Schroers and F.~A.~Bais,
\textit{``{Magnetic Charge Lattices, Moduli Spaces and Fusion Rules}''},
\textsf{\doiref{10.1016/j.nuclphysb.2008.08.003}{Nucl.~Phys.~B806,~386~(2009)}%
},
\texttt{\arxivref{0803.3376}}.
%
\bibitem{Kampmeijer:2008hw}
L.~Kampmeijer, F.~A.~Bais, B.~J.~Schroers and J.~K.~Slingerland,
\textit{``{Towards a non-abelian electric-magnetic symmetry: the skeleton
  group}''},
\textsf{JHEP~1001,~095~(2010)},
\texttt{\arxivref{0812.1230}}.
%
\bibitem{Fayet:1974jb}
P.~Fayet and J.~Iliopoulos,
\textit{``{Spontaneously Broken Supergauge Symmetries and Goldstone
  Spinors}''},
\textsf{\doiref{10.1016/0370-2693(74)90310-4}{Phys.~Lett.~B51,~461~(1974)}}.
%
\bibitem{Shifman:2007ce}
M.~Shifman and A.~Yung,
\textit{``{Supersymmetric Solitons and How They Help Us Understand Non-Abelian
  Gauge Theories}''},
\textsf{\doiref{10.1103/RevModPhys.79.1139}{Rev.~Mod.~Phys.~79,~1139~(2007)}},
\texttt{\arxivref{hep-th/0703267}}.
%
\bibitem{Isozumi:2004vg}
Y.~Isozumi, M.~Nitta, K.~Ohashi and N.~Sakai,
\textit{``{All exact solutions of a 1/4 Bogomol'nyi-Prasad- Sommerfield
  equation}''},
\textsf{\doiref{10.1103/PhysRevD.71.065018}{Phys.~Rev.~D71,~065018~(2005)}},
\texttt{\arxivref{hep-th/0405129}}.
%
\bibitem{Eto:2009bg}
M.~Eto et~al.,
\textit{``{Non-Abelian Vortices in SO(N) and USp(N) Gauge Theories}''},
\textsf{\doiref{10.1088/1126-6708/2009/06/004}{JHEP~0906,~004~(2009)}},
\texttt{\arxivref{0903.4471}}.
%
\bibitem{Auzzi:2003em}
R.~Auzzi, S.~Bolognesi, J.~Evslin and K.~Konishi,
\textit{``{Nonabelian monopoles and the vortices that confine them}''},
\textsf{\doiref{10.1016/j.nuclphysb.2004.03.003}{Nucl.~Phys.~B686,~119~(2004)}%
},
\texttt{\arxivref{hep-th/0312233}}.
%
\bibitem{Eto:2005yh}
M.~Eto, Y.~Isozumi, M.~Nitta, K.~Ohashi and N.~Sakai,
\textit{``{Moduli space of non-Abelian vortices}''},
\textsf{\doiref{10.1103/PhysRevLett.96.161601}{Phys.~Rev.~Lett.~96,~161601~(20%
06)}},
\texttt{\arxivref{hep-th/0511088}}.
%
\bibitem{Eto:2006uw}
M.~Eto, Y.~Isozumi, M.~Nitta, K.~Ohashi and N.~Sakai,
\textit{``{Manifestly supersymmetric effective Lagrangians on BPS solitons}''},
\textsf{\doiref{10.1103/PhysRevD.73.125008}{Phys.~Rev.~D73,~125008~(2006)}},
\texttt{\arxivref{hep-th/0602289}}.
%
\bibitem{Isozumi:2004jc}
Y.~Isozumi, M.~Nitta, K.~Ohashi and N.~Sakai,
\textit{``{Construction of non-Abelian walls and their complete moduli
  space}''},
\textsf{\doiref{10.1103/PhysRevLett.93.161601}{Phys.~Rev.~Lett.~93,~161601~(20%
04)}},
\texttt{\arxivref{hep-th/0404198}}.
%
\bibitem{Isozumi:2004va}
Y.~Isozumi, M.~Nitta, K.~Ohashi and N.~Sakai,
\textit{``{Non-Abelian walls in supersymmetric gauge theories}''},
\textsf{\doiref{10.1103/PhysRevD.70.125014}{Phys.~Rev.~D70,~125014~(2004)}},
\texttt{\arxivref{hep-th/0405194}}.
%
\bibitem{Eto:2004vy}
M.~Eto et~al.,
\textit{``{D-brane construction for non-Abelian walls}''},
\textsf{\doiref{10.1103/PhysRevD.71.125006}{Phys.~Rev.~D71,~125006~(2005)}},
\texttt{\arxivref{hep-th/0412024}}.
%
\bibitem{Eto:2005wf}
M.~Eto et~al.,
\textit{``{Global structure of moduli space for BPS walls}''},
\textsf{\doiref{10.1103/PhysRevD.71.105009}{Phys.~Rev.~D71,~105009~(2005)}},
\texttt{\arxivref{hep-th/0503033}}.
%
\bibitem{Eto:2006cx}
M.~Eto et~al.,
\textit{``{Non-Abelian vortices of higher winding numbers}''},
\textsf{\doiref{10.1103/PhysRevD.74.065021}{Phys.~Rev.~D74,~065021~(2006)}},
\texttt{\arxivref{hep-th/0607070}}.
%
\bibitem{Eto:2006db}
M.~Eto et~al.,
\textit{``{Universal reconnection of non-Abelian cosmic strings}''},
\textsf{\doiref{10.1103/PhysRevLett.98.091602}{Phys.~Rev.~Lett.~98,~091602~(20%
07)}},
\texttt{\arxivref{hep-th/0609214}}.
%
\bibitem{Eto:2006dx}
M.~Eto et~al.,
\textit{``{Non-Abelian duality from vortex moduli: a dual model of
  color-confinement}''},
\textsf{\doiref{10.1016/j.nuclphysb.2007.03.040}{Nucl.~Phys.~B780,~161~(2007)}%
},
\texttt{\arxivref{hep-th/0611313}}.
%
\bibitem{Eto:2007yv}
M.~Eto et~al.,
\textit{``{On the moduli space of semilocal strings and lumps}''},
\textsf{\doiref{10.1103/PhysRevD.76.105002}{Phys.~Rev.~D76,~105002~(2007)}},
\texttt{\arxivref{0704.2218}}.
%
\bibitem{Eto:2008yi}
M.~Eto et~al.,
\textit{``{Constructing Non-Abelian Vortices with Arbitrary Gauge Groups}''},
\textsf{\doiref{10.1016/j.physletb.2008.09.007}{Phys.~Lett.~B669,~98~(2008)}},
\texttt{\arxivref{0802.1020}}.
%
\bibitem{Eto:2009wq}
M.~Eto et~al.,
\textit{``{Multiple Layer Structure of Non-Abelian Vortex}''},
\textsf{\doiref{10.1016/j.physletb.2009.05.061}{Phys.~Lett.~B678,~254~(2009)}},
\texttt{\arxivref{0903.1518}}.
%
\bibitem{Eto:2010aj}
M.~Eto et~al.,
\textit{``{Group Theory of Non-Abelian Vortices}''},
\textsf{\doiref{10.1007/JHEP11(2010)042}{JHEP~1011,~042~(2010)}},
\texttt{\arxivref{1009.4794}}.
%
\bibitem{Fujimori:2010fk}
T.~Fujimori, G.~Marmorini, M.~Nitta, K.~Ohashi and N.~Sakai,
\textit{``{The Moduli Space Metric for Well-Separated Non-Abelian Vortices}''},
\textsf{\doiref{10.1103/PhysRevD.82.065005}{Phys.~Rev.~D82,~065005~(2010)}},
\texttt{\arxivref{1002.4580}}.
%
\bibitem{Eto:2004rz}
M.~Eto, Y.~Isozumi, M.~Nitta, K.~Ohashi and N.~Sakai,
\textit{``{Instantons in the Higgs phase}''},
\textsf{\doiref{10.1103/PhysRevD.72.025011}{Phys.~Rev.~D72,~025011~(2005)}},
\texttt{\arxivref{hep-th/0412048}}.
%
\bibitem{Eto:2005cp}
M.~Eto, Y.~Isozumi, M.~Nitta, K.~Ohashi and N.~Sakai,
\textit{``{Webs of walls}''},
\textsf{\doiref{10.1103/PhysRevD.72.085004}{Phys.~Rev.~D72,~085004~(2005)}},
\texttt{\arxivref{hep-th/0506135}}.
%
\bibitem{Eto:2005fm}
M.~Eto, Y.~Isozumi, M.~Nitta, K.~Ohashi and N.~Sakai,
\textit{``{Non-abelian webs of walls}''},
\textsf{\doiref{10.1016/j.physletb.2005.10.017}{Phys.~Lett.~B632,~384~(2006)}},
\texttt{\arxivref{hep-th/0508241}}.
%
\bibitem{Eto:2005sw}
M.~Eto, Y.~Isozumi, M.~Nitta and K.~Ohashi,
\textit{``{1/2, 1/4 and 1/8 BPS equations in SUSY Yang-Mills-Higgs systems:
  Field theoretical brane configurations}''},
\textsf{\doiref{10.1016/j.nuclphysb.2006.06.026}{Nucl.~Phys.~B752,~140~(2006)}%
},
\texttt{\arxivref{hep-th/0506257}}.
%
\bibitem{Eto:2006bb}
M.~Eto et~al.,
\textit{``{Effective action of domain wall networks}''},
\textsf{\doiref{10.1103/PhysRevD.75.045010}{Phys.~Rev.~D75,~045010~(2007)}},
\texttt{\arxivref{hep-th/0612003}}.
%
\bibitem{Eto:2007uc}
M.~Eto et~al.,
\textit{``{Dynamics of Domain Wall Networks}''},
\textsf{\doiref{10.1103/PhysRevD.76.125025}{Phys.~Rev.~D76,~125025~(2007)}},
\texttt{\arxivref{0707.3267}}.
%
\bibitem{MundetiRiera:1999fd}
I.~Mundet~i~Riera,
\textit{``{Yang-Mills-Higgs theory for symplectic fibrations}''},
\texttt{\arxivref{math/9912150}}.
%
\bibitem{Baptista:2004rk}
J.~M.~Baptista,
\textit{``{Vortex equations in abelian gauged sigma-models}''},
\textsf{\doiref{10.1007/s00220-005-1444-0}{Commun.~Math.~Phys.~261,~161~(2006)%
}},
\texttt{\arxivref{math/0411517}}.
%
\bibitem{Hindmarsh:1992yy}
M.~Hindmarsh,
\textit{``{Semilocal topological defects}''},
\textsf{\doiref{10.1016/0550-3213(93)90681-E}{Nucl.~Phys.~B392,~461~(1993)}},
\texttt{\arxivref{hep-ph/9206229}}.
%
\bibitem{Auzzi:2008wm}
R.~Auzzi, M.~Eto, S.~B.~Gudnason, K.~Konishi and W.~Vinci,
\textit{``{On the Stability of Non-Abelian Semi-local Vortices}''},
\textsf{\doiref{10.1016/j.nuclphysb.2008.12.024}{Nucl.~Phys.~B813,~484~(2009)}%
},
\texttt{\arxivref{0810.5679}}.
%
\bibitem{Auzzi:2007wj}
R.~Auzzi, M.~Eto and W.~Vinci,
\textit{``{Static Interactions of non-Abelian Vortices}''},
\textsf{\doiref{10.1088/1126-6708/2008/02/100}{JHEP~0802,~100~(2008)}},
\texttt{\arxivref{0711.0116}}.
%
\bibitem{Auzzi:2007iv}
R.~Auzzi, M.~Eto and W.~Vinci,
\textit{``{Type I Non-Abelian Superconductors in Supersymmetric Gauge
  Theories}''},
\textsf{\doiref{10.1088/1126-6708/2007/11/090}{JHEP~0711,~090~(2007)}},
\texttt{\arxivref{0709.1910}}.
%
\bibitem{Bowman:1985kh}
M.~C.~Bowman,
\textit{``{Parameter Counting For Selfdual Monopoles}''},
\textsf{\doiref{10.1103/PhysRevD.32.1569}{Phys.~Rev.~D32,~1569~(1985)}}.
%
\bibitem{Fujimori:2008ee}
T.~Fujimori, M.~Nitta, K.~Ohta, N.~Sakai and M.~Yamazaki,
\textit{``{Intersecting Solitons, Amoeba and Tropical Geometry}''},
\textsf{\doiref{10.1103/PhysRevD.78.105004}{Phys.~Rev.~D78,~105004~(2008)}},
\texttt{\arxivref{0805.1194}}.
%
\bibitem{Ferretti:2007rp}
L.~Ferretti, S.~B.~Gudnason and K.~Konishi,
\textit{``{Non-Abelian vortices and monopoles in SO(N) theories}''},
\textsf{\doiref{10.1016/j.nuclphysb.2007.07.021}{Nucl.~Phys.~B789,~84~(2008)}},
\texttt{\arxivref{0706.3854}}.
%
\bibitem{Eto:2008qw}
M.~Eto, T.~Fujimori, S.~B.~Gudnason, M.~Nitta and K.~Ohashi,
\textit{``{SO and USp K\'ahler and Hyper-K\'ahler Quotients and Lumps}''},
\texttt{\arxivref{0809.2014}}.
%
\bibitem{Gudnason:2010jq}
S.~B.~Gudnason and K.~Konishi,
\textit{``{Low-energy $U(1) \times USp(2M)$ gauge theory from simple high-
  energy gauge group}''},
\textsf{\doiref{10.1103/PhysRevD.81.105007}{Phys.~Rev.~D81,~105007~(2010)}},
\texttt{\arxivref{1002.0850}}.
%
\bibitem{Gudnason:2010rm}
S.~B.~Gudnason, Y.~Jiang and K.~Konishi,
\textit{``{Non-Abelian vortex dynamics: Effective world-sheet action}''},
\textsf{\doiref{10.1007/JHEP08(2010)012}{JHEP~1008,~012~(2010)}},
\texttt{\arxivref{1007.2116}}.
%
\bibitem{Atiyah:1984tk}
M.~F.~Atiyah,
\textit{``{Instantons In Two-Dimensions And Four-Dimensions}''},
\textsf{\doiref{10.1007/BF01212288}{Commun.~Math.~Phys.~93,~437~(1984)}}.
%
\bibitem{Nekrasov:2002qd}
N.~A.~Nekrasov,
\textit{``{Seiberg-Witten Prepotential From Instanton Counting}''},
\textsf{Adv.~Theor.~Math.~Phys.~7,~831~(2004)},
\texttt{\arxivref{hep-th/0206161}}.
%
\bibitem{Eto:2006mz}
M.~Eto et~al.,
\textit{``{Non-Abelian vortices on cylinder: Duality between vortices and
  walls}''},
\textsf{\doiref{10.1103/PhysRevD.73.085008}{Phys.~Rev.~D73,~085008~(2006)}},
\texttt{\arxivref{hep-th/0601181}}.
%
\bibitem{Eto:2007aw}
M.~Eto et~al.,
\textit{``{Statistical Mechanics of Vortices from D-branes and T- duality}''},
\textsf{\doiref{10.1016/j.nuclphysb.2007.06.020}{Nucl.~Phys.~B788,~120~(2008)}%
},
\texttt{\arxivref{hep-th/0703197}}.
%
\end{thebibliography}
\bibliographystyle{nb}

\end{document}